\documentclass[11pt, a4paper]{article}
\usepackage{jheppub} 

\pdfoutput=1 
\usepackage{graphicx}  
\usepackage{dcolumn}   
\usepackage{bm}        

\title{Statistical mechanics approach to lattice field theory}

\author[a]{Arturo Amador}
\author[a]{Johan S. H{\o}ye}
\author{K{\aa}re Olaussen}

\affiliation[a]{Institutt for fysikk, NTNU,\\N-7491 Trondheim}

\emailAdd{arturo.amador@ntnu.no}
\emailAdd{johan.hoye@ntnu.no}
\emailAdd{Kare.Olaussen@ntebb.no}

\abstract{The mean spherical approximation (MSA) is a closure relation for pair correlation functions
(two-point functions) in statistical physics. It can be applied to a wide range of systems, 
is computationally fairly inexpensive, and --- when properly applied and interpreted ---
lead to rather good results.

In this paper we promote its applicability to
euclidean quantum field theories formulated on a lattice, by demonstrating how
it can be used to locate the critical lines of a class of multi-component
bosonic models. The MSA has the potential to handle models lacking a positive
definite integration measure, which therefore are difficult to investigate by Monte-Carlo simulations.
}

\arxivnumber{1610.05284}
\date{\today}

\begin{document}

\maketitle






%


\section{Introduction}
\label{sec:Introduction}


\newcommand{\D}{N}
\newcommand{\m}{\,\texttt{m}}

Perturbative quantum field theories (QFT), most notably Quantum Electrodynamics,
belong to the most successful approaches in science. The agreement between
the predicted and experimental values of the electron magnetic moment is
probably the best verified number in physics. However, not all phenomena in this realm can
be reliably analysed by weak-coupling perturbation theory, which anyway has some inherent limitations, due to its
asymptotic nature and the fact that the expansion parameters are not always small.

It is not obvious how to proceed when perturbation theory fails, but one
well establish\-ed approach is to formulate an imaginary time version of
the relevant model on a lattice, and study this system by Monte-Carlo simulations.
Available computer memory and time impose
(steadily increasing) limits on the size of the systems that can be treated
by this method, and the accuracy of results.
But there are also interesting quantum field models which are difficult
to investigate by the Monte-Carlo method, due to lack of a
positive definite probability measure.

In this paper we discuss a different approach to lattice field theory, using methods which
have proven to work well in similar lattice models of statistical mechanics.
Ultimately they also require numerical work, e.g.~for the solution of integral equations.
We will mostly focus on the MSA, which is a fairly simple method for approximating
the pair correlation function (two-point function), and from this thermodynamic
properties of the system under scrutiny.

MSA is exact for gaussian models, e.g.~free
field theories, and (for the type of models we consider) to first order
of weak-coupling perturbation of these. It becomes exact for models
with $N$-component fields as $N\to\infty$, and becomes better when the number
of space-time dimensions $d$ increases. It is therefore expected to work better
for $d=4$ QFT models than the $d=3$ models of statistical mechanics
for which it was originally developed. It also becomes exact for
long range interactions, more precisely to first order of the
$\gamma$-expansion (although this is of less relevance for our
approach in this work). Being constrained by exact limits in
so many directions, the MSA have the prospects of being a quite relieable
and useful additional tool for analysis of QFT models.

\subsection{Summary of article}

The rest of this paper is organized as follows:
\begin{itemize}

\item[-]
In section~\ref{sec:FunctionalIntegral4QFT} we describe
the models to be analyzed. First in section~\ref{sec:ContinuumFI}
as a formal functional integral over functions $\bm{\varphi}(\bm{x})$ defined on a $d$-dimensional
euclidean space-time continuum. In section~\ref{sec:LatticeFI} this is
formulated as a related integral over variables $\bm{\varphi}_i$ defined on a discrete
$d$-dimensional hypercubic lattice; this can be interpreted as the
partition function for a classical lattice spin model, with $N$-dimensional
continuuous spins $\bm{s}_i$.

\item[-]
In section~\ref{sec:LatticeGasMixture} we reinterpret the lattice spin
model as describing a mixture of classical particles confined to the
sites of a lattice, with a hard-core interaction such that at most one
particle can be at each site. The value of $\bm{s}_i$ specifies the type
of particle at site $i$ (or if it is empty). In this interpretation the
interacting part of the system Hamiltonian consists the off-diagonal
terms of the lattice Laplace operator. In the absence of these terms
the model becomes \emph{ultra-local}; i.e.~it reduces to a product of
independent low-dimensional integrals, one for each lattice site. This
zero'th order model, usually denoted the \emph{reference system}, is
discussed in section~\ref{sec:ReferenceSystem}. Many quite successful
nonperturbative methods for treating interactions have been
developed for statistical mechanics of fluids. We describe some of
them in section~\ref{sec2.1}, including the $\gamma$-expansion, the
mean spherical approximation (MSA), and the self-consistent Ornstein-Zernike
approximation (SCOZA).

\item[-]
Free quantum field theories are described by gaussian functional
integrals, which therefore acts as the zero'th order model in
standard QFT perturbation expansions, while it is already
considered as an interacting model from our statistical mechanics
viewpoint. To contrast the two approaches we
therefore consider Gaussian models in
section~\ref{sec:GaussianModel}, for single-component
fields in section~\ref{subsec:SingleComponent},
and generalized to multi-component fields in
section~\ref{subsec:MultiComponent}.

\item[-]
For QFT applications without lattice artifacts, a lattice model
must be tuned to be very close to a second order critical point,
so that all local quantities of interest have essentially infinite
correlation lengths when measured
in lattice units. Thus, the first task of any lattice approach is to
locate the critical region. For the models considered explicitly
in this paper, defined by Euclidean actions of the form
\begin{equation}
	S = -\sum_{ij}\frac{1}{2} \bm{\varphi}_i \Delta^{(L)}_{ij} \bm{\varphi}_j
		+ \sum_i \frac{1}{2} m^2 \bm{\varphi}^2_i + 
		\frac{\lambda}{4!} \left(\bm{\varphi}^2_i\right)^2,
		\quad \lambda \ge 0,
\end{equation}
this consists of a line $m^2(\lambda)$ in the $(m^2, \lambda)$-plane,
with $m^2 \le 0$ starting at $m^2=0$ for $\lambda=0$. In
section~\ref{sec:PerturbationExpansion4CriticalLine} we calculate the
three first terms of a perturbation expansion of this line, using
standard weak-coupling QFT perturbation theory for $N$-component real fields.

\item[-]
In section~\ref{sec:MeanFieldCriticalLine} we use
spin model mean field theory (i.e., with averages computed in the
ultra-local model) to compute an estimate of the critical line.

\item[-]
In turns out that the results of the mean field calculations
are easily adapted to the corresponding MSA calculations,
performed in section~\ref{sec:MSAanalytic}, by simply
replacing the space-time dimension $d$ by an effective
dimension $d_{\text{eff}}$ (cf.~figure~\ref{fig:effectiveDimension}). Mathematically the two quantities
are related to the diagonal elements of the (inverse) lattice Laplace operators:
\begin{equation}
	1/(2d) = -\left[\Delta^{(L)}_{ii}\right]^{-1},\quad
	1/(2d_{\text{eff}}) = -\left[\Delta^{(L)}\right]^{-1}_{ii}\quad
	\text{(no sums over $i$)}.
\end{equation}
In section~\ref{sec:SmallLambdaExpansion} we compare the
weak coupling (small-$\lambda$) perturbation expansion
of the MSA-result with the exact expansion of
section~\ref{sec:PerturbationExpansion4CriticalLine}.
For the critical line, the MSA expansion is given by with
the same diagrams, but with the full lattice propagator
replaced by a constant. Contrary to the mean-field approximation
the MSA is exact to first order in $\lambda$.
 
In section~\ref{sec:LargeLambdaExpansion} we perform a
strong coupling (large-$\lambda$) perturbation expansion
of the MSA-result, and find a surprising ``duality'' relation between
coefficients of the weak and strong expansions,
cf.~eqs.~\eqref{strong2weakCorrespondence}. The
leading order of the strong coupling expansion corresponds
to the Ising limit, for which the $N=1$, $d=4$ critical point
is known from Monte-Carlo simulations. The MSA results differs
from this result by about $3.5\,\%$.

In section~\ref{sec:LargeNExpansion} we perform a $1/N$ expansion
of the MSA-result. The coefficients of this expansion exhibit
an curious ``duality'' (anti-)symmetry, originating from the
weak-strong relation mentioned above.

\item[-]
Section \ref{sec:MSAnumeric} describes some of the numerical computations
briefly and presents results for the critical line. Additional technical details are delegated to appendix A--D.

\end{itemize}

\section{Functional integral description of quantum field theories}
\label{sec:FunctionalIntegral4QFT}

In this section
we indicate briefly how bosonic quantum field theories (QFTs)
can be formulated on
a lattice, and interpreted as systems of continuous spins.
As we will see in section~\ref{sec:LatticeGasMixture}, such systems can in turn be viewed as models for classical particles
(lattice gases).

\subsection{Formal continuum description}
\label{sec:ContinuumFI}

In QFT the grand partition function for a system in thermal
equilibrium is formally given by
a functional integral of the form, 
\begin{equation}
\Xi = \text{e}^{\beta p V} = \int \text{e}^{-\frac{1}{\hbar c}\int^{\beta\hbar c}_0 \int_V d^dx\, \mathcal{L}(\bm{\varphi},\partial\bm{\varphi})}\;\mathcal{D}\bm{\varphi},
\label{QFTGrandPartitionFunction}
\end{equation}
where $\mathcal{L}$ is the Lagrangian of the QFT model, analytically
continued to imaginary time, $x^0 = \text{i}ct$, and $\beta = 1/k_B T$.
Here $V$ is the $(d-1)$-dimensional volume of space,
assumed taken to infinity in a regular manner.

For bosonic fields eq.~\eqref{QFTGrandPartitionFunction}
is of a form similar to the continuum
spin models we will discuss later (cf.
sections \ref{sec:LatticeGasMixture} and \ref{sec:GaussianModel}), but with
the important differences that 

\begin{itemize}

\item[(i)]
the temperature variable is related
to the extent of the imaginary time direction instead of being a parameter of the Lagrangian
$\mathcal{L}$,

\item[(ii)]
chemical potentials are related to constant external gauge fields
instead of local weight factors, $\partial/\partial x^0 \equiv 
\partial_0 \to \partial_0 - \mu/(\hbar c)$. One can
introduce an independent chemical potential for each independent
conserved current in the QFT model;
this has as consequence that the chemical potentials
of particles and anti-particles must
have opposite signs. This is a consequence of the fact
that particles can be freely created and
annihilated in relativistic QFT, only constrained
by the conservation laws of the model.

\item[(iii)]
the QFT mean energy is different from
the internal energy in the equivalent spin system.

\end{itemize}

The motivation/derivation of eq.~\eqref{QFTGrandPartitionFunction}
originates in the Feynman path integral formulation of quantum mechanics \cite{Feynman:1948ur, Feynman:1949zx},
and is exposed in depth in books
like \cite{pathFeynman, peskinqft, critical:props:kleinert}.
We give a brief indication of how it can be derived
in appendix \ref{GaussianModelFiniteTemperatureChemicalPotential},
with focus on the connection between the field theoretic and 
the conventional statistical mechanical description of the chemical
potential.

In spin models, or the related lattice gas models for classical particles,
the occurence of temperature and chemical potential parameters are
entirely different. One must beware of the confusions which
may arise when working and communicating from such disparate viewpoints.

In the remainder of this article we shall consider quantum systems at zero temperature and zero
chemical potential, specified by functional integrals of the form
\begin{equation}
\Xi_0\left\{\mathbf{h}_0\right\} = \int 
\text{e}^{-\int {d}^d x\, \left[-\frac{1}{2} \bm{\varphi}_0 \Delta \bm{\varphi}_0 + \frac{1}{2} m_0^2\,\bm{\varphi}_0^2 
	+ \frac{1}{4!} \lambda_0\,\left(\bm{\varphi}_0^2\right)^2 
	- \mathbf{h}_0\cdot \bm{\varphi}_0 \right]}\;\mathcal{D}\bm{\varphi}_0,
\label{PartitionFunction}
\end{equation}
where $\Delta$ is the Laplace operator in $d$-dimensional Euclidean space. We will mostly focus on the case of $d=4$.
In eq.~\eqref{PartitionFunction} the parameter $m_0^2$ (`bare mass')
may take negative values ---
which in fact is the most interesting case. The
field $\bm{\varphi}_0$ have $N$ components (often referred to
as \emph{spin dimension}). In some calculations we will restrict
$N$ to be a positive odd integer (even $N$ involves a
different set of special functions).

\subsection{Transformation to lattice formulation and spin model interpretation}
\label{sec:LatticeFI}

To make the model amenable to the statistical mechanical approach of
section~\ref{sec:LatticeGasMixture}, we restrict $x$ to the sites
$x_i$ of a hypercubic lattice,
with replacements $\int {d}^d x \to a^d \sum_{i}$ and  $\Delta \to a^{-2} \Delta^{(\text{L})}$.
Here $\Delta^{(\text{L})}$ is a corresponding dimensionless lattice Laplacian, with $a$ some characteristic
lattice length --- tiny relative to the scales of interest. By introducing new fields and parameters
\begin{equation}
\begin{split}
\bm{\varphi}_{i}    &= a^{(d-2)/2}\,\bm{\varphi}_0({x}_i),\quad
m^2         = a^2 m_0^2,\\
 \lambda   &=a^{(4-d)}\,\lambda_0,\quad
\mathbf{h}_i = a^{(d+2)/2}\,\mathbf{h}_0(x_i)
\end{split}
\end{equation}
we obtain a lattice partition function
\begin{equation}
\Xi\left\{\mathbf{h}\right\} = \int\,\text{e}^{\frac{1}{2}\sum_{ij}\bm{\varphi}_i \Delta^\text{(L)}_{ij} \bm{\varphi}_j 
	-\sum_{i}\left[\frac{1}{2} m^2 \bm{\varphi}_{i}^2 + \frac{1}{4!}\lambda {(\bm{\varphi}^2_{i})}^2 - \mathbf{h}_i\cdot \bm{\varphi}_i\right]}
\;\prod_{k} \frac{d \bm{\varphi}_{k}}{(2\pi)^{N/2}}.
\label{LatticePartitionFunction}
\end{equation}
This is similar to the partition functions
for classical spin models investigated in statistical mechanics,
most notably the Ising model, only in one dimension higher
than usual, and with continuous (perhaps multidimensional) spins.
We have a local factor for each site 
\begin{equation}
f(\bm{\varphi}_i)
= (2\pi)^{-N/2}\,\text{e}^{-\frac{1}{2}(m^2 + 2d)\bm{\varphi}_i^2 - \frac{1}{4!}\lambda (\bm{\varphi}_i^2)^2 + \mathbf{h}_i\cdot\bm{\varphi}_i} ,
\label{QFTChemicalPotential}
\end{equation}
plus nearest-neighbor ferromagnetic interactions 
$U = -\sum_{\langle i j\rangle} \bm{\varphi}_i\cdot \bm{\varphi}_j$, where
the sum runs over all nearest-neighbor pairs (``hopping Hamiltonian''),
corresponding to the standard numerical $(2d+1)$-stensil approximation of the
Laplace operator. 
By our statistical mechanical approach we will focus upon and utilize the MSA
(mean spherical approximation) which is defined and explained in section~\ref{sec:LatticeGasMixture}. 
For our MSA approach this particular form
of $U$ is not important;
the important feature is that
it can be expressed as a translation invariant sum
of quadratic pair interactions,
$U =  -\sum_{\langle ij \rangle} 
\bm{\varphi}_i\, \bm{\psi}(\mathbf{r}_{ij})\, \bm{\varphi}_j$.
Since a term of the form
$\frac{1}{2}\sum_{i} \bm{\varphi}_{i}\, \mathbf{c}_0\, \bm{\varphi}_{i}$ 
is both quadratic and local, we have the freedom to
multiply $f(\bm{\varphi}_i)$ by the factor
$\text{e}^{-\frac{1}{2} \bm{\varphi}_i\mathbf{c}_0\,\bm{\varphi}_i}$,
while changing 
$U \to U - \frac{1}{2}\sum_i \bm{\varphi}_i 
\mathbf{c}_0\,\bm{\varphi}_i$ at the same time. This freedom
is the most crucial part of the MSA; it resembles the procedure
of adding mass counterterms in standard
renormalized QFT perturbation theory. 

In the spin formulation the corresponding inverse temperature $\beta$
set to unity. This is different from the QFT $\beta$
of eq.~\eqref{QFTGrandPartitionFunction}.
For greater resemblance with statistical systems one may make a scale transformation of the fields,
$\bm{\varphi}_i = \sqrt{\beta}\, \bm{s}_i$, to eliminate one combination of the local parameters ($m^2$ and $\lambda$)
in favor of a temperature-like variable. This is useful to simplify
comparison with known results in the Ising limit, $\bm{s}^2_i = 1$.

\section{Lattice gas mixtures}
\label{sec:LatticeGasMixture}

In this section we discuss the connection
between spin models (also known as bosonic QFTs) and
multicomponent mixtures of classical particles. It is well known that the Ising model can be regarded as a lattice gas, i.e. a gas of classical particles confined to the sites $i$ of a lattice, where Ising spin $s_{i}=1$ defines the presence of a particle on site $i$,
and $s_{i}=-1$ that site $i$ is empty.  Note that $i$ may label the sites of a multidimensional lattice; hence it is naturally viewed as an index \emph{vector},
but to simplify notation we do not write this explicitly.
The Ising model can be extended to spins taking  more that two values, and eventually to continuous spins. 
Such systems can be regarded as mixtures of classical particles confined to the sites of a lattice (i.e., lattice gas mixtures). The advantage of employing this equivalence 
is that physical intuition, together with some powerful
methods developed for analysis of classical fluids,
can be applied fruitfully. Such an approach was
used by H\o ye and Stell in a previous study
of continuous spins \cite{hoye97}.
Here we will use it to study continuous spins in four spatial dimensions.

In the lattice gas interpretation the density of particles of type $s$ at a given site $i$ is
\begin{equation}
	\rho_s = \langle n_s \rangle,
\end{equation}
where the set of particle numbers at that site, $\left\{ n_s  \right\}$,  is restricted by the hard core condition, $\sum_{s\ne0} n_s \le 1$,  preventing multiple occupancy of cells.
Here an empty site (vacuum state) is specified by $n_0 = 1$.  Therefore, the configuration at site $i$ is uniquely specified
by the value of $s$ for which  $n_s = 1$. With discrete spins the $s$ takes discrete values, but we will eventually
let $s$ be a continuous variable. 

Consider a lattice gas mixture in $d$ dimensions, 
where the lattice is taken to be simple cubic. The particles of type $s$ have a chemical potential $\mu_s$. 
With the equivalent spin systems in mind,
the {fugacity} 
in general can be written as
\begin{equation}
	z_s=e^{\beta\mu_s}=f(s)\,
	\text{e}^{\beta{\cal H}s},
\label{1}
\end{equation}
where ${\cal H}$ is the applied magnetic field, $\beta=1/(k_B T)$, $T$ is temperature, and $k_B$  is Boltzmann's constant. 
By specifying $z_0 = f(0) =1$, the vacuum state is normalized to have zero  ``chemical potential'', $\mu_0=0$.

In the lattice gas interpretation the (grand) partition function for this system is identified by the thermodynamic pressure $p$,
\(
       Z_{g} = \text{e}^{\mathcal{N} \beta p}
\),
where $\mathcal{N}$ is the number of lattice sites. We associate each site with a cell of unit volume. In this representation
the thermodynamic potential $\beta p$ is viewed as a function of the chemical potentials $\beta \mu_s$. For 
later statistical mechanical evaluations it is convenient to make a Legendre transformation to new independent
variables and a new thermodynamic potential,
\begin{subequations}
\begin{align}
      \rho_s &= \frac{\partial(\beta p)}{\partial(\beta\mu_s)},\\
      I         &= \beta p - \sum_s\beta \mu_s \rho_s = \Big(1 - \sum_s \beta\mu_s \frac{\partial}{\partial(\beta \mu_s)} \Big)\beta p =  -\beta F.
\label{1b}
\end{align}
\end{subequations}
Here $F = U -TS$ is the Helmholtz free energy
per site, where
$U$ is the (configurational) internal energy
per site and $S$ is the corresponding entropy.

\subsection{The reference system}
\label{sec:ReferenceSystem}

We will consider expansions around a zero'th order system, the reference system, where there is no interaction
between particles at different sites.
This is different from the free field (gaussian) models commonly used as zero'th
order systems in QFTs. We will denote quantities related to the reference system
by the subscript $R$. 
The grand partition function per site is simply
\begin{equation}
	Z_g= \text{e}^{\beta p_R} = \mathop{{\sum}'}_{\left\{ n_s\ge 0\right\} } \text{e}^{\beta \mu_s n_s}  = \sum_s z_s.
\label{3}
\end{equation}
Here the first sum is restricted by the hard core condition, $\sum_{s\ne 0 } n_s \le 1$.
This leads to the second sum. 
In the limit of continous spins the second sum
should be replaced by a corresponding integral over $s$.
The average particle densities become
\begin{equation}
      \rho_s = \frac{\partial\ln Z_g}{\partial(\beta\mu_s)}=\frac{z_s}{Z_g}.
\label{4}
\end{equation}
From this follows the pressure $p_R$ of the reference system hard core lattice gas,
\begin{equation}
	\beta p_R=\ln Z_g=\ln(z_0/\rho_0)= -\ln \rho_0=-\ln(1-\rho),
\label{5}
\end{equation}
with total density $\rho=\sum_{s\neq0} \rho_s$, and $\rho_0$ the density of empty sites. At low total density this reduces
to the ideal gas law. For a Legendre transformation we use eqs.~\eqref{4}--\eqref{5} to express
the chemical potentials in terms of densities,
\begin{subequations}
\begin{equation}
          \beta\mu_s =        \ln\frac{\rho_s}{\rho_0},
\label{5a}
\end{equation}
which gives
\begin{equation}
	I_R \equiv -\beta F_R=-\sum_{s\neq 0}\rho_s\ln \rho_s-\rho_0\ln\rho_0.
\label{6}
\end{equation}
\end{subequations}
From these results one may verify the thermodynamic relations,
\begin{subequations}
\begin{align}
	\beta\mu_s &= -\frac{\partial I_R}{\partial \rho_s}, \label{7}\\
	G_R &=\sum_{s\neq 0}\rho_s\mu_s=F_R +p_R,
\label{7a}
\end{align}
\end{subequations}
where $G_R$ is the Gibbs free energy per site for the reference system.
 
For the expansion in section \ref{sec2.1}, some correlation functions of the reference system are also needed.
Since there are no interactions between different sites,
these functions are ultra-local. The quantities of interest
are\footnote{The letter `m' is a standard symbol for both particle mass
and magnetization $\langle s \rangle$. To distinguish we use $m$
for particle mass, and ${\m}$ for magnetization.}
\begin{subequations}
\begin{align}
	\sum_s\,\langle n_s \rangle s &= \sum_s \rho_s\,s  =  \langle s \rangle \equiv {\m},\label{MeanValue}\\
	\sum_{ss'} \langle n_{s} n_{s'} \rangle\, s s' &= \sum_{ss'} \rho_s \delta_{ss'} \,ss' = \langle s^2 \rangle, \label{Correlator}
\end{align}
or more precisely in the combination
\begin{equation}
       R \equiv \langle  s^2 \rangle - {\m}^2.
\label{7c}
\end{equation}
\end{subequations}
The first equality in eq.~\eqref{Correlator} follows from the hard core condition, which implies that $ n_s n_{s'} = n_s \delta_{ss'}$. 
 
\subsection{Interacting systems and the $\gamma$-expansion}
\label{sec2.1}

The spins or lattice gas particles may interact. The usual form of this interaction is a sum of pair interactions of the Heisenberg form,
\begin{equation}
   U= \frac{1}{2}\sum_{ij}  \sum_{s_i,s_j} n_{s_i}(i)\, n_{s_j}(j)\, \psi({\bf r}_{ij})s_i s_{j}
 = \frac{1}{2}\sum_{i,j}\psi({\bf r}_{ij})s_i s_{j},
\label{9a}
\end{equation}
with $\psi(\mathbf{0})=0$, and where $\psi<0$ for ferromagnetic interactions.
Here ${\bf r}_{ij}={\bf r}_i-{\bf r}_j$ is the relative separation between spins $i$ and $j$. The hard core interaction that prevents multiple occupancy
is similar to the hard cores of molecules in classical fluids. When the interaction (\ref{9a}) is turned on, the $s$-dependence of the density distribution (\ref{4}) will change. For the Ising model there is no such change for a given magnetization,
since it corresponds to a one component lattice gas (with only two possible states at each site). 
However, for the more general situation a satisfactory approximation of the resulting $s$-dependence is crucial.

Similar to classical fluids, the solution of the spin problem with non-zero interactions (\ref{9a}) cannot be done
exactly (with a few exceptions, mostly in one and two dimensions). 
Thus, one has to make approximations, usually described by a sum of graphs. E.g.~to go beyond a pure density expansion,
some organizing principle for summing classes of graphs is required. One such principle is
the $\gamma$-ordering scheme by Hemmer \cite{Hemmer64} and Lebowitz et.~al.~\cite{Lebowitz65}, where one
makes the replacement
\begin{equation*}
       \psi(\textbf{r}_{ij}) \rightarrow \gamma^d\,\psi(\gamma \textbf{r}_{ij})
\end{equation*}
in eq.~\eqref{9a}, continues with a systematic expansion around $\gamma=0$, and (for a concrete physical system) sets $\gamma=1$ in the end.
The resulting organization of diagrams has close similarity to the loop (or $\hbar$) expansion in quantum field theory. 
The leading contribution is the mean field approximation,  which becomes exact in the limit $\gamma\rightarrow 0$.

In terms of the statistical mechanical graph expansion of the free energy as function of densities,
the whole mean field contribution is given by a single potential bond, i.e.
\begin{equation}
\begin{split}
  I_{MF} &= I_R + \frac{1}{2\mathcal{N}} \sum_{i,j} \sum_{s_i,s_j} \rho_{s_i} \rho_{s_j}v(\mathbf{r}_{ij})s_i s_j\\
&= I_R + \frac{1}{2} {\m}^2\,\tilde v(\mathbf{0}).
\label{9b}
\end{split}
\end{equation}

We have assumed the densities $\rho_s$ to be independent of
position.
The tilde designates a $d$-dimensional Fourier transformed function,
\begin{equation}
\begin{split}
\tilde v(\mathbf{k}) &= \sum_j v(\mathbf{r}_{0j})\,\text{e}^{\text{i}\mathbf{k}\cdot\mathbf{r}_{0j}},\\
v(\mathbf{r}) &= \frac{1}{(2\pi)^d}\int_{-\pi}^{\pi}\cdots\int_{-\pi}^{\pi}\,\tilde v(\mathbf{k})\,\text{e}^{\text{i}\mathbf{k}\cdot\mathbf{r}} d\mathbf{k}.
\end{split}
\end{equation}
\begin{equation}
\tilde v(\mathbf{k}) = c_0-\beta\tilde\psi({\bf k}). \label{10}
\end{equation}
An adjustable parameter, $v(\mathbf{0}) \equiv c_0$ (with $\psi(\mathbf{0})=0$), has now been introduced. 
The $c_0$ corresponds to a finite particle-particle interaction inside the hard cores ($\mathbf{r}= \mathbf{0}$). 
This does not influence the exact physics of the system, since two particles cannot be at the same site.

\begin{figure}[h!]
\begin{center}
\includegraphics[clip, trim=0ex 0ex 0ex 0ex,width=0.8 \textwidth]{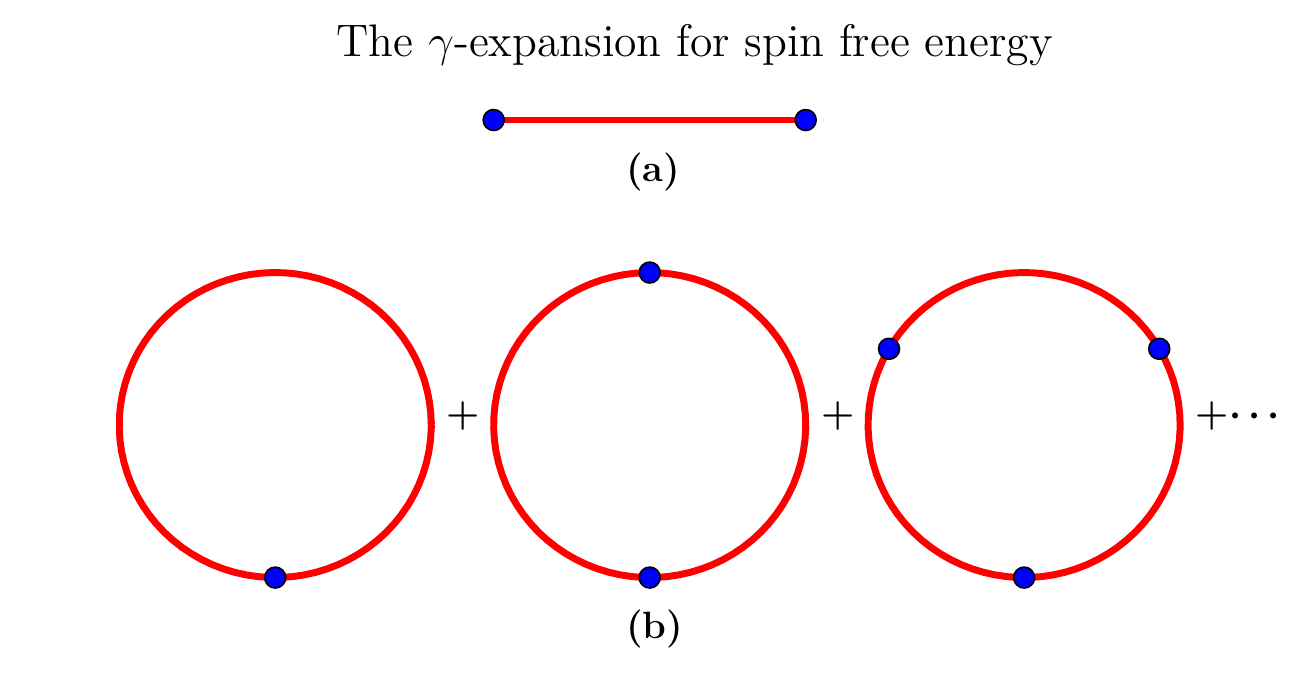}  
\caption{\label{fig:gammaExpansion} 
Diagrams for the first two terms of the $\gamma$-expansion. 
A potential bond (red line)
represents the function $v(\textbf{r}_{ij})\,s_{i} s_{j} \equiv [c_0\delta_{ij} - \beta\psi(\textbf{r}_{ij})]\,s_{i} s_{j}$ in real space,
and each vertex (filled, blue circle) represents a summation $\sum_{s} \sum_{s'} s\tilde\Gamma_{0ss'} s'$ where $s$ and $s'$ are the endpoints of potential bonds on each side of the vertex.
}
\end{center}
\end{figure}

The first correction to the mean field approximation is given by the ring graphs, see figure \ref{fig:gammaExpansion}(b),
whose sum is
\begin{equation}
	I_{M1}=-\frac{1}{2(2\pi)^d}\int_{-\pi}^\pi\cdots\int_{-\pi}^\pi\,
	\ln(1-R\tilde v(\mathbf{k}))\,d\mathbf{k}
	\label{9}.
\end{equation}
However, with the particle picture one has to show a bit care here since $c_0$ varies. Then a potential bond should not return to the same $\rho_s$ vertex. This means that the $\rho_s\delta_{ss'}$ and $\langle s^2\rangle$ terms of eqs.~(\ref{11}) and (\ref{12}) below should not contribute in the first ring graph (with one blue circle) of figure~\ref{fig:gammaExpansion}. Thus this should be compensated by which the particle picture contribution to the free energy is ($\psi(\mathbf{0})=0$)
\begin{eqnarray}
I_1&=&I_{M1}-\frac{1}{2}c_0\langle s^2 \rangle= 
I_{M1}-\frac{1}{2}c_0(R+{\m}^2)\nonumber \\
&=&I_{M1}-\frac{1}{2(2\pi)^3}\int R\tilde v(\mathbf{k})\,d\mathbf{k}
-\frac{1}{2}c_0 {\m}^2.
\label{10a}
\end{eqnarray}
This result is also consistent with the thermodynamics of the MSA obtained in ref.~\cite{hoye77}. The result for $I_1$ is the same as for graph expansion with $c_0$ constant. A reason for this is that first partial derivatives with respect to $c_0$ (and thus $R$) will cancel as seen below where the equation of state is found via the chemical potentials $\mu_s$.

The constant $R$ in eqs.~(\ref{9}) and (\ref{10a}) represents local (hyper-)vertices in the ring graphs, where pairs of potential bonds meet.
The $s$-dependences of their endpoints are summed (averaged) over the reference system pair correlation functions
that also includes self-correlations ($\rho$-vertex). They are
\begin{equation}
\tilde\Gamma_{0ss'}({\bf k})=\Gamma_{0ss'}(0)=\rho_s\delta_{ss'}-\rho_s\rho_{s'}
\label{11}
\end{equation}
since $\Gamma_{0ss'}({\bf r})=0$ otherwise. The last term of this equation expresses the hard core condition that prohibits more than one particle to occupy a cell (i.e.~zero minus the ideal gas probability). With this one gets
\begin{equation}
\begin{split}
   R&=\sum_{ss'}ss'\Gamma_{0ss'}(0)=\langle s^2 \rangle-{\m}^2,\\
   {\m}&=\langle s \rangle=\sum_s s \rho_s,
\label{12}
\end{split}
\end{equation}
where ${\m}$ is average magnetization.

Now we can find the leading contribution to the resulting pair correlation function with $U\neq 0$. It is formed by chain graphs where $\tilde v({\bf k})$-bonds are connected by hypervertices. Its average when including the reference piece (\ref{12}) is
\begin{equation}
\tilde\Gamma({\bf k})=\frac{R}{1-R\tilde v({\bf k})}.
\label{13}
\end{equation}

We can now optimize the value of the parameter $c_0$ such that the exact core condition expressed by eq.~(\ref{11}) also is fulfilled by the resulting correlation function. With the average (\ref{12}) this implies the condition
\begin{equation}
R=\Gamma(\mathbf{0})=\frac{1}{(2\pi)^d}\int\tilde\Gamma({\bf k})\,d{\bf k}.
\label{14}
\end{equation}
Due to this condition the resulting contribution to the chemical potential from the $I_1$ of eq.~(\ref{10a}) simplifies; so it becomes
\begin{equation}
\beta\mu_{1s}=-\frac{\partial I_1}{\partial\rho_s}=-c_0 {\m}s+\frac{1}{2}c_0 s^2.
\label{15}
\end{equation}
This follows from $\partial R/\partial\rho_s=s^2-2{\m}s$ with $R$
given by eq.~(\ref{12}). Further, differentiation of $c_0$ does not
contribute due to condition (\ref{14}) \cite{hoye77}.
The same is the situation for $R$ except for the last
term of eq.~(\ref{9}) which results in
$\tilde v(\mathbf{k})\rightarrow v(\mathbf{0})=c_0-\beta\psi(\mathbf{0})=c_0$.
Altogether, including result (\ref{15}), the mean field contribution
from eq.~(\ref{9b}), and the reference system contribution~(\ref{5a})
the resulting chemical potentials become
\begin{equation}
\beta\mu_s=\ln\left(\frac{\rho_s}{\rho_0}\right)-
[c_0-\beta\tilde\psi(\mathbf{0})]{\m}s+\frac{1}{2}c_0s^2.
\label{16}
\end{equation}
When inserted in eq.~(\ref{1}) one finds the resulting density or effective spin distribution 
\begin{equation}
z_{es}=\frac{\rho_s}{\rho_0}=f_e(s) e^{\beta {\cal H}_e s}
\label{17}
\end{equation}
with
\begin{equation}
\beta{\cal H}_e =\beta{\cal H}+[c_0-\beta\tilde\psi(0)]{\m},
\quad f_e(s)=f(s)e^{-c_0 s^2/2}.
\label{18}
\end{equation}
This equation also gives the equation of state ${\cal H}={\cal H}({\m})$
where ${\cal H}_e={\cal H}_e({\m})$ is the equation of state for
free spins that follows from eq.~(\ref{17}) alone. 

The above expressions with $c_0$ determined via the exact core condition as expressed by eq.~(\ref{14}) is the MSA (mean spherical approximation) extended to mixtures where $\langle s^2\rangle$ is not fixed. Its equation of state as a magnetic spin system is in the MSA given by eq.~(\ref{18}). 

The MSA, which is consistent with $\gamma$-expansion, has its origin in the SM (spherical model) that was solved by Berlin and Kac as an approximation to the Ising  model \cite{berlin52}. In the SM the values of the spins are not restricted to $\pm1$; instead the sum of their values squared are fixed. This was modified by Lewis and Wannier to the MSM (mean SM) where the average of the spin values squared are fixed \cite{lewis52}. This is equivalent to a Gaussian model with adjustable one-particle potential to keep $\langle s_i^2\rangle=1$ fixed. Then the MSM was extended to continuum fluids by Lebowitz and Percus \cite{lebowitz66}. This extension was the basis of the MSA with correlation function eq.~(\ref{13}) and core condition eq.~(\ref{14}) in the present case.
With the MSA the reference system is defined
to be the one of non-interacting Ising spins.
The MSM is different in this respect.
In section \ref{sec2.1} the MSA is extended
to a more general spin system.
(However, since we in this work only need to
consider ${\m}=0$, the difference from MSA is
not significant.)

From eq.~(\ref{18}) one can evaluate the inverse susceptibility $\chi^{-1}$
\begin{equation}
\beta\chi^{-1}=\frac{\partial{\cal H}}{\partial {\m}}
=\frac{1}{R}-\tilde v(\mathbf{0})-\frac{\partial c_0}{\partial {\m}}{\m}
\label{19}
\end{equation}
with $\partial(\beta{\cal H}_e)/\partial{\m}=1/R$.
According to the fluctuation theorem one should
have $\beta\chi^{-1}=\Gamma(0)^{-1}$. Expression (\ref{19})
deviates from this by the term {\,} $-(\partial c_0/\partial {\m}){\m}$.
Such deviations are typical for any approximation
and is some measure of its accuracy. This inaccuracy
can be removed by the SCOZA (self-consistent
Ornstein-Zernike approximation) where a free
parameter like effective temperature is
used \cite{hoye84,hoye85}. The latter has lead
to very accurate results for instance for the Ising model
in one, two, and three dimensions. Accurate numerical
SCOZA data for the Ising model was initially
obtained by Dickman and Stell \cite{dickman96}.
With nearest neighbor interaction it is exact
in one dimension. In two dimensions the sharp
phase transition is missing, but otherwise
the results are very accurate when compared
with the known exact solution for ${\m}=0$ \cite{hoye98}. 

Later a variety of accurate SCOZA results have been obtained.
Various references to such results can be found in the work
by H{\o}ye and Lomba in their detailed investigation of
the critical properties of another related accurate
theory, the HRT (hierarchical reference theory) (in three
spatial dimensions) \cite{hoye11}. They found that the
critical indices turned out to be simple rational numbers.
Recently this analysis of the HRT was extended to spins
of dimensionality\footnote{In the works referred to here the symbol $D$
is commonly used for spin dimensionality.} ${\D}$, and they found by analysis in
view of SCOZA and detailed numerical work that critical
indices were independent of ${\D}$ \cite{lomba14}. This
contrasts earlier HRT results and other previous results
from renormalization group theory where indices are
expected to vary with ${\D}$ \cite{parola12,fisher72}.
The reason for rational numbers and independence
upon ${\D}$ (for ${\D}$ finite) of the critical indices
is the found connection between a leading and two
subleading layers of contributions to the critical
behavior. These layers are connected to each other
and mean field behavior away from the critical
point by which rational numbers for the critical
indices appear. (However, the indices with good
accuracy may effectively vary with ${\D}$ when results
are fitted to an assumed single power.) \cite{lomba14}.

The HRT, based upon the momentum space renormalization
group theory, developed by Wilson and Kogut \cite{wilson74},
was introduced and studied by Parola and Reatto \cite{parola85}.
However, the SCOZA method, which was extended to
continuous spins in ref.~\cite{hoye97},
will not be pursued further in this work.

\section{The Gaussian model}
\label{sec:GaussianModel}

It may be instructive see how the method of
section \ref{sec:LatticeGasMixture} works
on an explicitly solvable example, the Gaussian model.
Essentially, this model is just a different name
for a free bosonic lattice field theory (allowing
for a more general Hamiltonian),
but viewed and analysed from the perspective of
statistical mechanics. The partition function
and all correlation functions
are in principle straightforward to calculate by
doing gaussian integrals. In this section we want
to reconstruct these results.
In section \ref{subsec:SingleComponent} we discuss
this model for the case of a one-dimensional continuous
spin $s$ (single-component field),
as in section \ref{sec:LatticeGasMixture}.
In section \ref{subsec:MultiComponent} we
generalize to an $N$-dimensional continuous spin
(multi-component field).

\subsection{Single-component field (one-dimensional spin)} 
\label{subsec:SingleComponent}

This model is defined by the quadratic form
\begin{equation}
\beta H = \frac{1}{2} \sum_{i, j} {s}_i \mathcal{K}(\textbf{r}_{ij}) {s}_{j} - \beta\mathcal{H} \sum_i s_i,
\label{Hamiltonian0}
\end{equation}
where each $s_i$ take continuous values in the range  $\left(-\infty,\infty \right)$,
and 
\begin{equation}
\mathcal{K}_{ij} \equiv \mathcal{K}(\textbf{r}_{ij}) = R_0^{-1} \delta_{ij} +\beta\psi(\textbf{r}_{ij})
\end{equation}	
is a positive definite matrix which we have decomposed into a local term, $R_0^{-1} \delta_{ij}$, and an interaction term, $\psi_{ij} \equiv \psi(\textbf{r}_{ij})$ with $\psi(\mathbf{0})=0$.
The partition function and lowest order correlation functions for a number of $\mathcal{N}$ sites evaluates to
(with $\int \mathcal{D}s \equiv \prod_k \int_{-\infty}^{\infty} ds_k/\sqrt{2\pi}$) ($\psi_{ij}\rightarrow \psi$ with matrix $\psi$),
\begin{subequations}
	\label{DirectAnswer}
	\begin{align}
	\text{e}^{\mathcal{N} \beta p} &= \int \mathcal{D}s\,\text{e}^{-\frac{1}{2} \sum_{ij} s_i \left(R_0^{-1} +\beta\psi\right)_{ij} s_j 
		+ \beta\mathcal{H} \sum_i s_i}\nonumber\\ 
	&=   \det{}^{-1/2} \left(R_0^{-1} +\beta\psi \right)\, \text{e}^{\frac{1}{2} \mathcal{N} 
		\left[R_0^{-1} +\beta\tilde{\psi}(\mathbf{0})\right]^{-1}\, (\beta\mathcal{H})^2}\nonumber\\ 
	&= \det{}^{-1/2} \left(R_0^{-1} +\beta\psi\right)\, \text{e}^{\frac{1}{2} \mathcal{N} 
		\left[R_0^{-1} +\beta \tilde{\psi}(\mathbf{0})\right]\, m^2},\label{Pressure}\\
	{\m} &\equiv \langle s_i \rangle =  \big[{R_0^{-1} +\beta\tilde{\psi}(\mathbf{0})}\big]^{-1}\,\beta \mathcal{H}, 
	\label{3.3b}\\
	\Gamma(\textbf{r}_{ij}) &\equiv \langle s_i s_j \rangle - \langle s_i \rangle \langle s_j \rangle =  \big({R_0^{-1} +\beta\tilde{\psi}(\mathbf{0})}\big)^{-1}_{ij}.
	\label{3.3c}
	\end{align}
\end{subequations}
The contribution from the determinant in eq.~\eqref{Pressure} can be written as
\begin{align}
&\log \det{}^{-1/2}\left(R_0^{-1} +\beta\psi\right) \nonumber\\ 
&= \frac{\mathcal{N}}{2} \log R_0 - 
\frac{\mathcal{N}}{2(2\pi)^d} \int_{-\pi}^\pi \cdots\int_{-\pi}^\pi \log\left[1 - R_0(-\beta \tilde{\psi})(\mathbf{k})\right]\,d\bf{k}. 
\end{align}
All higher order correlators can be constructed from ${\m}$ and $\Gamma$ by use of Wick's theorem. Further, all subsets of the $s$-variables are gaussian distributed, 
with parameters which can be found from the expressions above. 
In particular, the $s$-distribution at each single site is found to be
\begin{equation}
\rho_s = \frac{1}{\sqrt{\Gamma(\mathbf{0})}}\,
\text{e}^{-\frac{1}{2} (s-{\m})^2/\Gamma(\mathbf{0})}.
\label{3.5}
\end{equation}
when the ``normalization'' $\sum_s\rho_s\rightarrow\int\rho_s\,ds/\sqrt{2\pi}$ like the one in eq.~(\ref{Pressure}) is used.

Then we will test the lattice gas method on the Gaussian model. The given spin distribution will be $f(s)=e^{-s^2/(2R_0)}$. 
With the MSA the effective spin distribution and equation of state are given by eqs.~(\ref{17}) and (\ref{18}). Thus $f_e(s)=\exp[-(R_0^{-1}+c_0)s^2/2]$, and accordingly
\begin{equation}
	{R}^{-1}= {R_0}^{-1}+c_0.
\label{26}
\end{equation}
By insertion of expression (\ref{26}) into eq.~(\ref{13})
the exact correlation function (\ref{3.3c}) is recovered.
Further by computing the equation for $\beta{\cal H}_e({\m})$
for the reference system from $f_e(s)$ one finds
\begin{equation} 
\beta{\cal H}_e={\m}/R={\m}/R_0+c_0 {\m} 
\label{26a}
\end{equation}
(i.e.~eq.~(\ref{3.3b}) with $R_0\rightarrow R$ and $\psi=0$).
When inserted in eq.~(\ref{18}) the exact result (\ref{3.3b})
for ${\m}$ is recovered. With spin distribution (\ref{17})
the exact one (\ref{3.5}) is also recovered when $R=\Gamma(\mathbf{0})$
from eq.~(\ref{14}) is used.

The gaussiam model partition function can also be found by use of the MSA. This is done in Appendix \ref{secA} where the exact result (\ref{Pressure}) is recovered.

Altogether we have found that the MSA solves the gaussian
model exactly for any pair interaction (\ref{9a}).
For other situations it becomes exact in the
mean field limit $\gamma\rightarrow 0$ where
$\gamma$ is the inverse range of interaction.
Finally it becomes exact in the limit $N\rightarrow\infty$
for spins of dimensionality $N$ \cite{YanWannier65, hoye1997Ddim}.
The latter is equivalent to the MSM (mean spherical model)
which has been generalized to the GMSM (generalized MSM) \cite{hoye07}.



\subsection{Multi-component field ($N$-dimensional spins)}
\label{subsec:MultiComponent}

The generalization to multi-component gaussian fields is mostly a change of notation. The quadratic form becomes
\begin{equation}
\beta H = \frac{1}{2} \sum_{i, j}\bm{s}_{i}\big[ \mathbf{R}_0^{-1}\delta_{ij} +\beta \mathbf{\psi}(\bm{r}_{ij}) \big] \bm{s}_{j}
- \beta\bm{\mathcal H}\cdot  \sum_i \bm{s}_i,
\label{Hamiltonian1}
\end{equation}
where each $\bm{s}_i$ and $\bm{\mathcal{H}}$ are $N$-component real vectors, 
$\mathbf{R}$ is a real symmetric positive definite $N\times N$ matrix, and $\mathbf{v}(\bm{r})$
is a real matrix-valued function (symmetric $N\times N$ matrices). With 
$\int (\cdots) \mathcal{D}\bm{s} \equiv \prod_k  \int_{-\infty}^{\infty} (\cdots) {d\bm{s}_k}/{({2\pi})^{N/2}}$
the partition function, and associated correlation functions, evaluates to
\begin{subequations}
	\begin{align}
	\text{e}^{\mathcal{N}\beta p} &=  \int
	\text{e}^{-\frac{1}{2}\sum_{ij}\bm{s}_i\left[\mathbf{R}_0^{-1} +\beta \mathbf{\psi} \right]_{ij} 
		\bm{s}_j +\beta \bm{\mathcal{H}}\cdot\sum_i\bm{s}_i} \; \mathcal{D}\bm{s} \nonumber\\
	&= \det{}^{-1/2}(\mathbf{R}_0^{-1} +\beta\mathbf{\psi})\,
	\text{e}^{\frac{1}{2} \mathcal{N} (\beta\bm{\mathcal{H}})[\mathbf{R}_0^{-1}+\beta\mathbf{\tilde{\psi}}(\bm{0})]^{-1} (\beta\bm{\mathcal{H}})}\nonumber\\
	&= \det{}^{-1/2}(\mathbf{R}_0^{-1} +\beta\mathbf{\psi})\,
	\text{e}^{\frac{1}{2} \mathcal{N} \bm{m}\,[\mathbf{R}_0^{-1}+\beta\mathbf{\tilde{\psi}}(\bm{0})]\, \bm{m}},\\
	\bm{m} &= \langle \bm{s}_i \rangle = \big[ \mathbf{R}_0^{-1} +\beta \mathbf{\tilde{\psi}}(\bm{0}) \big]^{-1} \beta\bm{\mathcal{H}},\\
	\bm{\Gamma}(\mathbf{r}_{ij}) &=  \langle \bm{s}_i \bm{s}_j \rangle - 
	\langle\bm{s}_i \rangle \langle\bm{s}_j \rangle = \left[ \mathbf{R}_0^{-1} +\beta \mathbf{\psi} \right]^{-1}_{ij}. 
	\end{align}
\end{subequations}
The $\bm{s}$-distribution at each single site is similar to eq.~(\ref{3.5})
\begin{equation}
\rho_{\bm{s}} = \det \big[\bm{\Gamma}(\mathbf{0})\big]^{-1/2}\,
\text{e}^{-\frac{1}{2} (\bm{s}-\bm{m}) {\bm{\Gamma}(\bm{0})}^{-1} (\bm{s}-\bm{m})}.
\end{equation}
As we did for the single-component gaussian model above, one can again perform the MSA evaluations to recover the exact solution.

\section{Perturbation expansion for the critical line}
\label{sec:PerturbationExpansion4CriticalLine}

For later comparison, we shall in this section compute the critical line for small values of $\lambda$ by
the renormalized field theoretic perturbation method.
The requirement of infinitely long-range correlations is that the renormalized mass $m^2_r$ vanishes,
\begin{equation}
	m^2_r = m^2 + \delta m^2(\lambda) = 0.
	\label{RenormalizedMass}
\end{equation}
Hence, we use $-\frac{1}{2}\sum_{ij}\bm{\varphi}_i \Delta^\text{(L)}_{ij} \bm{\varphi}_j $ as the $0^\text{th}$
order Hamiltonian (we do not use a renormalized
field $\bm{\varphi}$). This gives a $0^\text{th}$ order lattice propagator,
\begin{align}
   \mathbf{G}^{(0)}_{ab}(\bm{x}; m^2) &= \delta_{ab}\,\frac{1}{V_\text{BZ}}
   \int_{\text{BZ}} \frac{\text{e}^{\text{i}\bm{k}\cdot\bm{x}}}{m^2 + 2d - 2 \sum_{n=1}^d \cos(k_n)}\; \text{d}\bm{k}
   \nonumber\\&=
   \delta_{ab}\,\int_0^\infty \text{e}^{-m^2 t}\,\prod_{n=1}^d \text{e}^{-2t}\,I_{x_n}(2t)\,\text{d}t, 
\end{align}
where the last integral follows from appendix \ref{LatticePropagator}.

\begin{figure}[h!]
\begin{center}
\includegraphics[clip, trim=15ex 108ex 15ex 26ex,
width=0.9 \textwidth]{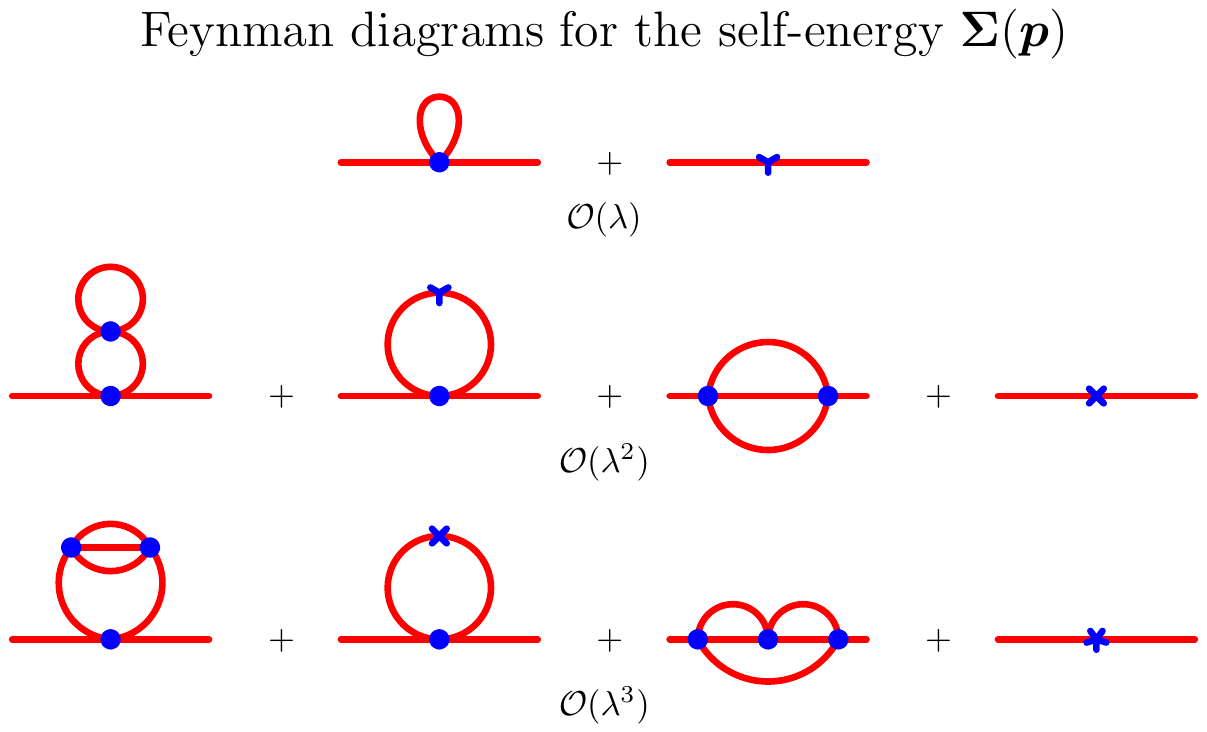}
\end{center}
\caption{\label{fig:SelfEnergyDiagrams}Self-energy diagrams to third order in $\lambda$. There is an
overall minus-sign, and a factor $-\lambda$ for each ordinary vertex.
The $m^2$-contributions
(last term of each line) is chosen such that the contributions of each line vanishes at $\bm{p} = \bm{0}$. Due to this
condition the first two terms on the $\mathcal{O}(\lambda^2)$-line cancel exactly. There is a similar cancelation between tadpole terms of
order $\lambda^3$; we have not drawn the cancelling diagrams on the $\mathcal{O}(\lambda^3)$-line. For $N=1$ the combinatorial factors
are given directly by the symmetry factors of each diagram. For larger $N$ one may rewrite 
$\frac{\lambda}{4!}(\bm{\varphi}^2)^2 = \frac{\lambda}{4!}\sum_a \varphi^4_a + \frac{\lambda'}{2!^3}\sum_{a\ne b}\varphi^2_a \varphi^2_b$,
with $\lambda'=\frac{1}{3}\lambda$,  and use the symmetry factors of a larger set of diagrams with the same topology, but propagators
of different colors.
}
\end{figure}

Next compute the self-energy correction $\mathbf{\tilde{\Sigma}}(\bm{p}; \lambda)$ to
$\mathbf{G}(\bm{x}_{ij}; m^2, \lambda) \equiv \langle \bm{\varphi}_i \bm{\varphi}_j \rangle$ in powers of $\lambda$,
and choose $m^2$ order by order in $\lambda$ such that eq.~\eqref{RenormalizedMass} is fulfilled. The
Feynman diagrams for $\mathbf{\tilde{\Sigma}}(\bm{p})$ are shown
in figure~\ref{fig:SelfEnergyDiagrams}.
This determines the critical values of $m^2$ to
\begin{subequations}
\label{PerturbationExpansionOfCriticalLine}
\begin{align}
	m^2 = &-{\textstyle \frac{(N+2)}{6}}\lambda\,G^{(0)}(\bm{0}; 0) + 
	{\textstyle \frac{(N+2)}{18}} \lambda^2 \sum_i {G^{(0)}(\bm{x}_i;0)}^3\nonumber\\
	&-{\textstyle \frac{(N+2)^2}{108}} \lambda^3 \sum_{i} G^{(0)}(\bm{x}_i;0)^3 
	\sum_j G^{(0)}(\bm{x}_j;0) \left[ G^{(0)}(\bm{x}_j + \bm{x}_i;0) - G^{(0)}(\bm{x}_j;0)\right]\nonumber\\
	&-{\textstyle \frac{(N+2)(N+8)}{108}} \lambda^3 \sum_i G^{(0)}(\bm{x}_i;0)^2 \sum_j  G^{(0)}(\bm{x}_j; 0)^2\, G^{(0)}(\bm{x}_i + \bm{x}_j;0)
	\label{DescriptiveFormula}\\
	= &-{\textstyle \frac{(N+2)}{6}}\lambda\, a_1 + {\textstyle \frac{(N+2)}{18}}\lambda^2\, a_2 - 
	{\textstyle \frac{(N+2)^2}{108}}\lambda^3\,a_{3a} -
	{\textstyle \frac{(N+2)(N+8)}{108}}\lambda^3\,a_{3b}.\label{QFTWeakCouplingExpansion}
\end{align}
\end{subequations}
The terms in eq.~\eqref{DescriptiveFormula} are combined such that all sums are convergent.
Numerical evaluation of the sums (rather crude for the higher order diagrams) gives
\begin{align*}
	a_1     &=  \phantom{-}0.154\,933\,390\,231\,060\,214\,084\,837\,208\ldots\\
	a_2     &=  \phantom{-}0.004\,043\,054\,812\,2\ldots\\
	a_{3a} &= -0.000\,007\,513\ldots\\
	a_{3b} &=  \phantom{-}0.000\,109\,381\ldots
\end{align*}

\section{Mean field calculation of the critical line}
\label{sec:MeanFieldCriticalLine}

Consider the field with local factor $f(\varphi_i)$
given by eq.~(\ref{QFTChemicalPotential}) and nearest
neighbor interaction $U$ as described below this equation.
(Note here the subscript $i$ denotes position in space.)
The vector notation means $N$-component fields which
we will consider for ${\bf h}_i={h}=0$. 

Assume that $\bm{\varphi}$ develops a vacuum expectation value, and orient coordinates
such that $\langle \bm{\varphi}_a \rangle =  \bar{\varphi}\, \delta_{a1}$. In mean field theory
$\bar{\varphi}$ is determined self-consistently from the relation (with $\beta\tilde\psi(\mathbf{0})=-2d$)
\begin{equation}
       \bar{\varphi} = \frac{\int \bm{\varphi}_1\,\text{e}^{-\frac{1}{2}\left(m^2 + 2d\right)\,\bm{\varphi}^2
       - \frac{1}{4!}\lambda (\bm{\varphi}^2)^2 + 2d\, \bar{\varphi} \bm{\varphi}_1}\, \text{d}\bm{\varphi}}{
       \int \text{e}^{-\frac{1}{2}\left(m^2 + 2d\right)\,\bm{\varphi}^2 - \frac{1}{4!}\lambda\,(\bm{\varphi}^2)^2 
       + 2d\, \bar{\varphi} \bm{\varphi}_1} \text{d}\bm{\varphi}}.
	\label{MeanFieldCondition}
\end{equation}
With a second order transition the critical point is given by the limit $\bar\varphi \to 0$.
In this limit the condition~\eqref{MeanFieldCondition} becomes
\begin{equation}
	\frac{N}{2d} = \langle \bm{\varphi}^2 \rangle \equiv
	\frac{\int \bm{\varphi}^2\,\text{e}^{-\frac{1}{2}\left(m^2 + 2d\right)\,\bm{\varphi}^2
       - \frac{1}{4!}\lambda (\bm{\varphi}^2)^2}\, \text{d}\bm{\varphi}}{
       \int \text{e}^{-\frac{1}{2}\left(m^2 + 2d\right)\,\bm{\varphi}^2 - \frac{1}{4!}\lambda\,(\bm{\varphi}^2)^2} \text{d}\bm{\varphi}}.
       \label{MeanFieldCriticalLine}
\end{equation}
The right hand side of eq.~\eqref{MeanFieldCriticalLine} does not depend on dimension $d$. The integrals can be expressed
by Bessel functions, cf.~appendix~\ref{subsec:ExactExpressions}.

\section{MSA calculation of the critical line}
\label{sec:MSAanalytic}

Critical points are located where the pair correlation function becomes long- ranged, i.e. where its Fourier transform diverges. Unless periodic ordering is present this divergence takes place at $\mathbf{k} = 0$. Thus the transition is located 
where the denominator of expression (\ref{13}) vanishes
\begin{equation}
1-R\tilde v(\mathbf{0}) = 0 \label{48notes}
\end{equation}
with $\tilde{v}(\mathbf{k}) = c_0 - \beta \tilde{\psi}(\mathbf{k})$
and $R = \langle s^2\rangle-{\m}^2$. For the symmetric case with the
transition in zero magnetic field one has ${\m} = \langle s \rangle = 0$.
With given $\beta \tilde{\psi}(\mathbf{k})$, the $R$ and $c_0$ are
determined from the core condition (\ref{14}) ($R = \Gamma(\mathbf{0})$)
and the effective MSA spin distribution (\ref{18}) such that (${\m} = 0$)

\begin{figure}[h!]
\begin{center}
\includegraphics[width=0.9 \textwidth]{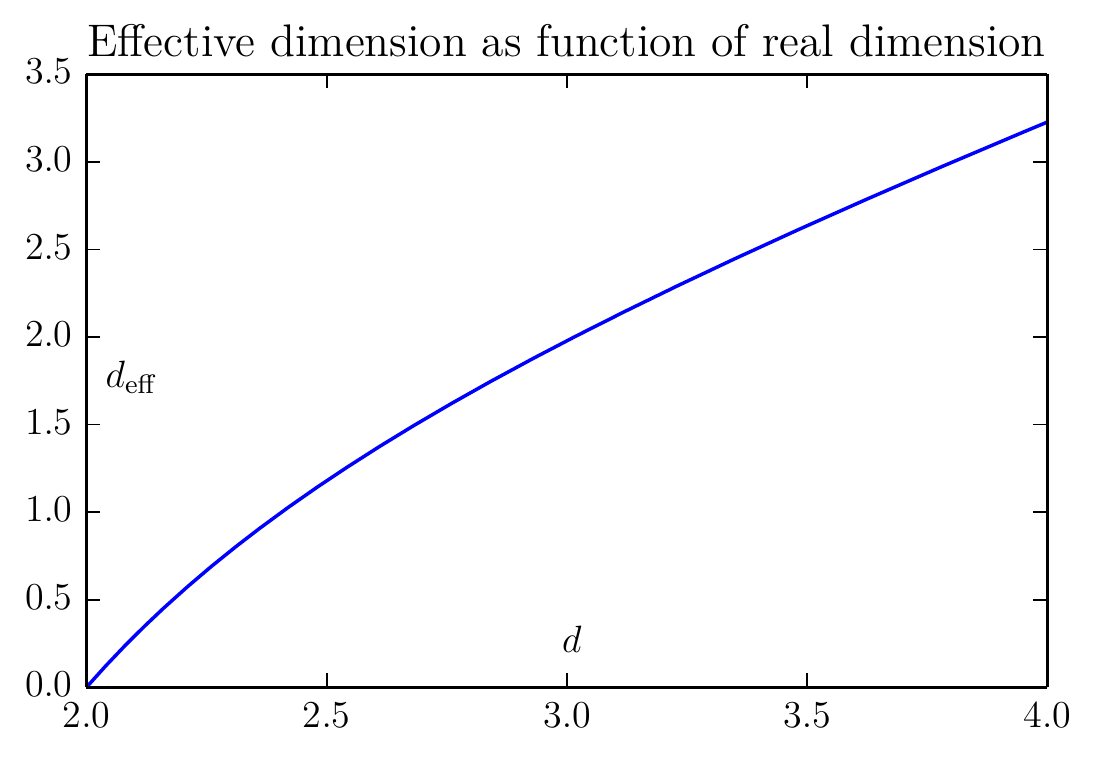}
\end{center}
\caption{\label{fig:effectiveDimension}Effective dimension
$d_{\text{eff}}=1/(2R)$ from the MSA solution that follows
from the integral~(\ref{53notes}). Note that $R$ is independent of spin dimensionality.
}
\end{figure}

Hence the MSA condition for the critical line becomes
\begin{equation}
	R = \langle s^ 2 \rangle = \frac{\int s^2 f(s)
		\text{e}^{-c_0 s^2/2} \ ds}{\int f(s) e^{-c_0 s^2/2} \ ds}.
	    \label{MSACriticalLine}
\end{equation}
For multicomponent spins the $s$ becomes a vector $\mathbf{s}$, and the $R$ is replaced with $RN$, i.e. $R = \langle s_1^2 \rangle$ ($N$ is spin dimension). 

For the field theory considered in the mean field limit in section 6
we have the equivalent spin model with
$s \rightarrow \varphi \rightarrow \boldsymbol{\varphi}$.
The corresponding spin distribution is
\begin{equation}
	f(\boldsymbol{\varphi}) = e^{-\frac{1}{2}(m^2 + 2d)
	\boldsymbol{\varphi}^2 -
	\frac{1}{4!} \lambda(\boldsymbol{\varphi}^2)^2}
	\label{50notes}
\end{equation}
with nearest neighbor interaction such that
\begin{equation}
	\beta \tilde{\psi}(\mathbf{0}) = -2d.
	\label{51notes}
\end{equation}

When the inverse range of attraction $\gamma \rightarrow 0$
the mean field limit is obtained. In the limit the
$\tilde{\psi}(\mathbf{k})$ will approach zero except for
a peak of width $\propto \gamma$ around $\mathbf{k} = 0$.
Thus in the limit $\gamma \rightarrow 0$
the $\tilde{\psi}(\mathbf{k})$ will no longer contribute
to the core condition integral (\ref{14}) by which the
parameter $c_0$ will be zero. Thus the effective spin
distribution (\ref{18}) will be the given one eq.~(\ref{50notes}).
With eqs.~(\ref{13}) and (\ref{51notes}) this at the critical
point implies
\begin{equation}
R = \frac{1}{2d}.
\label{52notes}
\end{equation}
Altogether, with $NR = \int \boldsymbol{\varphi}^2 f(\boldsymbol{\varphi})\ d\boldsymbol{\varphi}/\int f(\boldsymbol{\varphi})\ d\boldsymbol{\varphi}$,
eq.~(\ref{MeanFieldCriticalLine}) is recovered in the limit $\gamma \rightarrow 0$.

With finite $\gamma$, however, the $c_0$ will be non-zero,
and the core condition integral can be written as
\begin{align}
 R &= \frac{1}{(2\pi)^d} \int \frac{R_e \ d\mathbf{k}}{1 - R_e(-\beta \tilde{\psi}(\mathbf{k}))}
 = \frac{1}{2d}\frac{1}{(2\pi)^d} \int \frac{d\mathbf{k}}{1 - (-\beta \tilde{\phi}(\mathbf{k})/2d)},
\label{53notes}
\end{align}
with $R_e = R/(1-Rc_0)$ where $R_e=1/(2d)$ at the critical point. From this one finds $c_0 = R^{-1} - R_e^{-1} = R^{-1} - 2d$. With (\ref{18}) this means that the effective spin distribution becomes expression (\ref{50notes}) with $2d$ replaced by $R^{-1}$. Hence the MSA condition for the critical line becomes
\begin{equation}
	\text{Tr}\,\mathbf{G}^{(0)}(\bm{0}) = N R 
	=\frac{\int \bm{\varphi}^2\,\text{e}^{-\frac{1}{2}\left(m^2 + R^{-1}\right)\,\bm{\varphi}^2
       - \frac{1}{4!}\lambda (\bm{\varphi}^2)^2}\, \text{d}\bm{\varphi}}{
       \int \text{e}^{-\frac{1}{2}\left(m^2 + R^{-1}\right)\,\bm{\varphi}^2 - \frac{1}{4!}\lambda\,(\bm{\varphi}^2)^2} \text{d}\bm{\varphi}}.
       \label{MSACriticalLine2}
\end{equation}
This is just the mean field condition \eqref{MeanFieldCriticalLine} with $(2d)^{-1}$ replaced by 
$R = \int_0^\infty \text{e}^{-2dt}\,I_0(2t)^d\, dt$
which follows when eqs.~(\ref{c1}) and (\ref{c2})
for nearest neighbor interaction are applied to
eq.~(\ref{53notes}). Altogether, as far as the
critical line is concerned, the MSA method is equivalent
to mean field theory in an effective dimension
$d_\text{eff} = \frac{1}{2} R^{-1}$ (independent of $N$),
as plotted in figure~\ref{fig:effectiveDimension}. 

Series solutions of eq.~\eqref{MSACriticalLine2} with respect
to $m^2$ can be found for small and large values of $\lambda$,
and are given in the following sections.

\subsection{Small-$\lambda$ expansion of the MSA solution}
\label{sec:SmallLambdaExpansion}

For small $\lambda$ we find, cf.~appendix~\ref{subsec:SmallLambda},  
\begin{align}
	R m^2 = &- (N+2){\textstyle\big( \frac{R^2 \lambda}{6}\big)} + 
	2 (N+2)\left[ {\textstyle \big(\frac{R^2 \lambda}{6}\big)^2} 
	- (N+8)  
	{\textstyle \big(\frac{R^2 \lambda}{6}\big)^3}\right. \nonumber\\ &+
	(N^2 + 26 N + 108){\textstyle \big(\frac{R^2\lambda}{6}\big)^4} - 
	(N^3+58N^2+684 N + 1\,984){\textstyle \big(\frac{R^2\lambda}{6}\big)^5} \label{SmallLambdaExpansion} \\&+\left. 
	(N^4 + 108 N^3 + 2\,584 N^2 + 19\,824 N + 45\,280)
	{\textstyle \big(\frac{R^2\lambda}{6}\big)^6} + \cdots\right].\nonumber
\end{align}
This expansion has the same form as eq.~\eqref{PerturbationExpansionOfCriticalLine}, but with no sum over lattice points
(i.e.~like a zero-dimensional model),
with the propagator replaced by the constant $R=a_1$. Hence, the MSA reproduces first order of perturbation
theory exactly. It further predicts a magnitude of the second order term ($a_2$) which is about 8\% too small, and
of the third order term which (for $N=1$) is about 20\% too small.

\subsection{Large-$\lambda$ expansion of the MSA solution}
\label{sec:LargeLambdaExpansion}

For large $\lambda$ we find, cf.~appendix~\ref{subsec:LargeLambda},
\begin{align}
    N R m^2 =  &-{\textstyle \frac{N^2 R^2\lambda}{6}} -2  + 2(N-2)\left[{\textstyle \big( \frac{6}{N^2 R^2 \lambda}\big)}
    - (N-8) {\textstyle \big( \frac{6}{N^2 R^2 \lambda}\big)^2}    
    \right.\nonumber\\
    &+(N^2 -26N+108){\textstyle \big( \frac{6}{N^2 R^2 \lambda}\big)^3}
    - (N^3-58 N^2 + 684 N -1\,984)
    {\textstyle \big( \frac{6}{N^2 R^2 \lambda}\big)^4}\label{LargeLambdaExpansion}\\ 
    &+ \left.
    (N^4 - 108 N^3 + 2\,584 N^2 - 19\,824 N + 45\,280)
    {\textstyle \big( \frac{6}{N^2 R^2 \lambda}\big)^5} + \cdots\right].\nonumber
\end{align}
This expression may seem to indicate an exact result
for $N=2$. This is not the case; we see from
eq.~\eqref{Large_Lambda_Expansion} that there are contributions
which vanishes exponentially fast as $\lambda\to\infty$.
Such terms contributes to the small-$\lambda$ perturbation expansion
in eq.~\eqref{SmallLambdaExpansion}. I.e., the large-$\lambda$ expansion
to all orders --- even if it could be summed to an exact expression ---
may not reproduce the small-$\lambda$ behavior (and \emph{vice versa}).
In reality, the expansion constitutes only an asymptotic series, in
which case it becomes an interesting question whether the infinite
set of expansion coefficients contain sufficient information to
reproduce the exact result.

Surprisingly, a comparison of eqs.~\eqref{SmallLambdaExpansion} and
\eqref{LargeLambdaExpansion} reveals that the coefficients
of the large-$\lambda$ expansion, except the leading one, are intriguingly related to
the coefficients of the small-$\lambda$ expansion.
Specifically, for the terms inside the brackets, by the relations
\begin{subequations}
\label{strong2weakCorrespondence}
\begin{align}
	\big({\textstyle\frac{6}{N^2 R^2\lambda}}\big)^k
		&\rightarrow (-1)^{k+1} 
		\big(\textstyle{\frac{R^2 \lambda}{6}}\big)^{k+1},
		\label{Large2Small}\\
	N &\rightarrow -N, \label{NegativeDimension}
\end{align}
\end{subequations}
for $k=1,2,\ldots$. We have not uncovered the origin of this duality
relation; it is reason to expect that it is related to a
Hubbard-Stratanovich transformation of the integrals
in eq.~\eqref{MSACriticalLine2}.

The Ising limit is obtained as $\lambda\to\infty$, with $m^2 = -\frac{1}{6} R\lambda$ on the critical line. 
We introduce a new variable $s = \varphi/\sqrt{R}$ to impose the Ising spin condition, $s^2 = 1$.
This scaling introduces a temperature parameter $\beta = R$ in the interaction term. Hence
the MSA predicts a critical temperature $\beta_c =R =  0.154\,933\ldots$ for the $4d$ Ising model.
This model has been investigated by Gaunt, Sykes and McKenzie \cite{GauntSykesMcKenzie79}
using series expansion methods, and more recently by Lundow and 
Markstr{\"o}m \cite{LundowMarkstrom09} using Monte-Carlo simulations,
indicating a critical temperatature of
$\beta_c = 0.149\,694\,7 \pm 5\cdot 10^{-7}$.
I.e., the MSA prediction is only $3.5\%$ too large in the Ising limit.
Since the MSA is exact to first order in $\lambda$,
and also becomes exact in the limit $N\to\infty$,
the Ising limit ($\lambda\to\infty$, $N=1$) is probably the most inaccurate case of
the models we have analyzed, since it is the opposite of these
exact limits.

With SCOZA, where thermodynamic self-consistency is imposed, accuracy is expected to increase further, as discussed a bit 
at the end of section \ref{sec2.1}.

\subsection{$1/N$-expansion of the MSA solution}
\label{sec:LargeNExpansion}
%

We introduce
$u = \frac{1}{6} N {R}^2\,\lambda$ and consider the limit where $N$ becomes
large with $u$ fixed. This defines a $1/N$-expansion, cf. appendix \ref{subsec:LargeN},
\begin{align}
	R\, m^2 =&-u + \sum_{k\ge1} c_k(u)\,N^{-k}\nonumber\\
	=&-u
	-\frac{1}{1+u}\,{\left(\frac{2u}{N} \right)}
	+\frac{(1-u)}{(1 + u)^4}\,{\left(\frac{2u}{N} \right)^2}
	-\frac{4(1-3u+u^2)}{(1+u)^7}\,{\left(\frac{2u}{N} \right)^3}\nonumber\\
	&+\frac{(1-u)(27-122u+27u^2)}{(1 + u)^{10}}\,{\left(\frac{2u}{N} \right)^4}
	\label{OneOverNexpansion}
	\\&
	-\frac{4\,(62 - 521 u + 990 u^2 - 521 u^3 + 62 u^4)}{(1+u)^{13}}\,
	{\left(\frac{2u}{N} \right)^5} + \cdots\nonumber
\end{align}

\begin{figure}
\includegraphics[width=0.9 \textwidth]{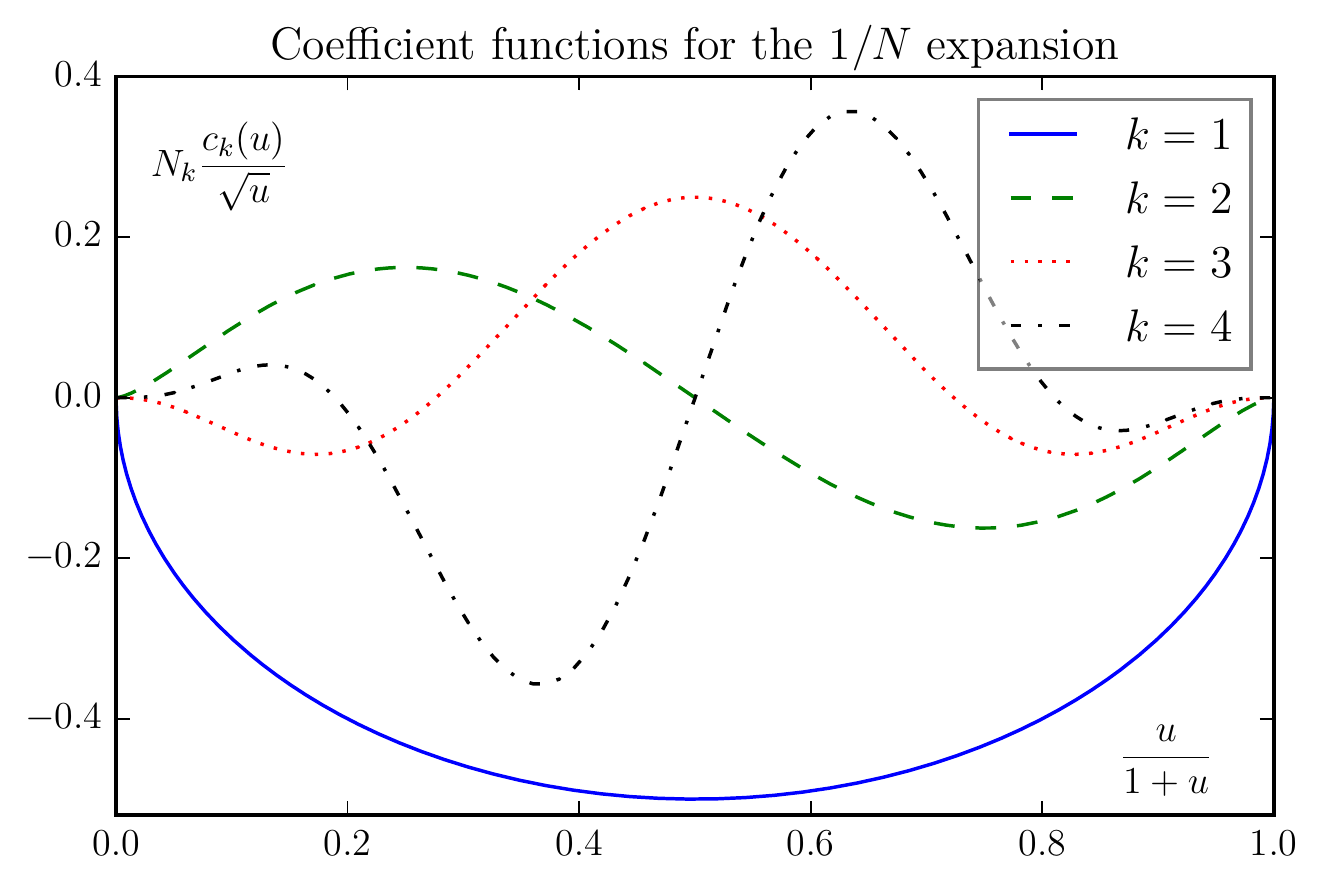}
\caption{\label{fig:oneOverNcoeffs}
The first few coefficient functions of the $1/N$-expansion \eqref{OneOverNexpansion}.
The parameter $x \equiv u/(1+u)$ used on the $x$-axis maps to $1-x$ when $u\to 1/u$.
The quantities $c_k(u)/\sqrt{u}$ used on the $y$-axis
maps to $(-1)^{k-1}\,c_{k}(u)/\sqrt{u}$. The constant $N_k$ equals $\frac{1}{2}$ for $k=1$,
otherwise 1.
}
\end{figure}

A further expansion of this expression in powers of $u$ agrees with a
$1/N$-expansion of eq.~\eqref{SmallLambdaExpansion}. A further expansion of this
expression in powers of $u^{-1}$ agrees with the $1/N$-expansion
of eq.~\eqref{LargeLambdaExpansion}. But also the expansion \eqref{OneOverNexpansion} is inexact, since
we leave out exponentially small (in $N-2$) contributions
to the integrals in eq.~\eqref{largeN_MSA}.
Note again the curious $u\to 1/u$ (anti-)symmetry of the coefficient functions present
in eq.~\eqref{OneOverNexpansion}, which is revealed explicitly in
figure \ref{fig:oneOverNcoeffs}.

\begin{figure}
\begin{center}
\includegraphics[width= 0.9 \textwidth]{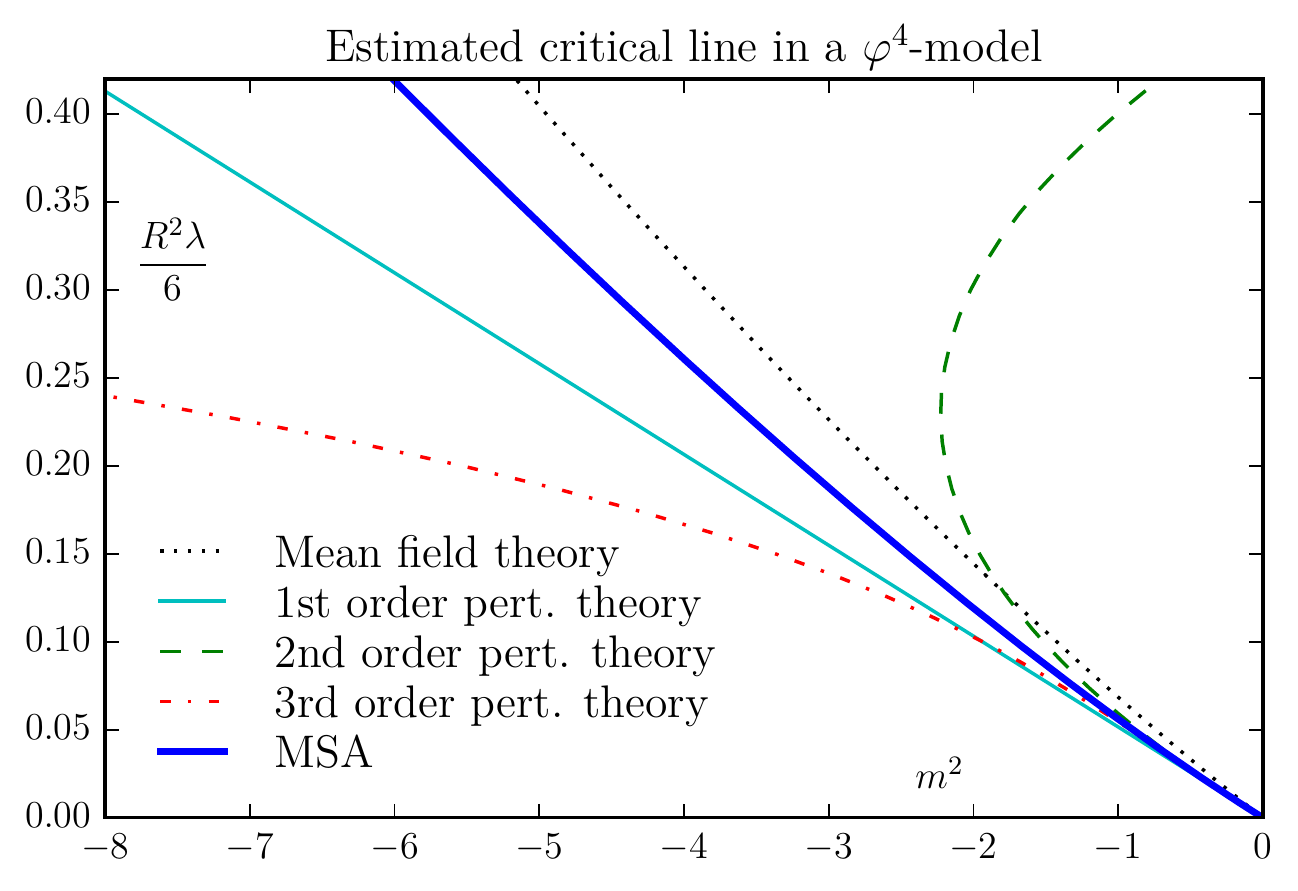}
\end{center}
\caption{\label{figure5}Some predictions of the critical
line in the 4-dimensional $\varphi^4$-model (for $N=1$).
The mean field result is inaccurate already to first order
in $\lambda$. Visually, the MSA prediction is almost
indistinguishable from the perturbation expansion
in the region where the latter looks trustable.
To better expose the differences we present the same
results in a different way in figure~\ref{figure6}.
}
\end{figure}

\begin{figure}
\begin{center}
\includegraphics[width=0.9 \textwidth]{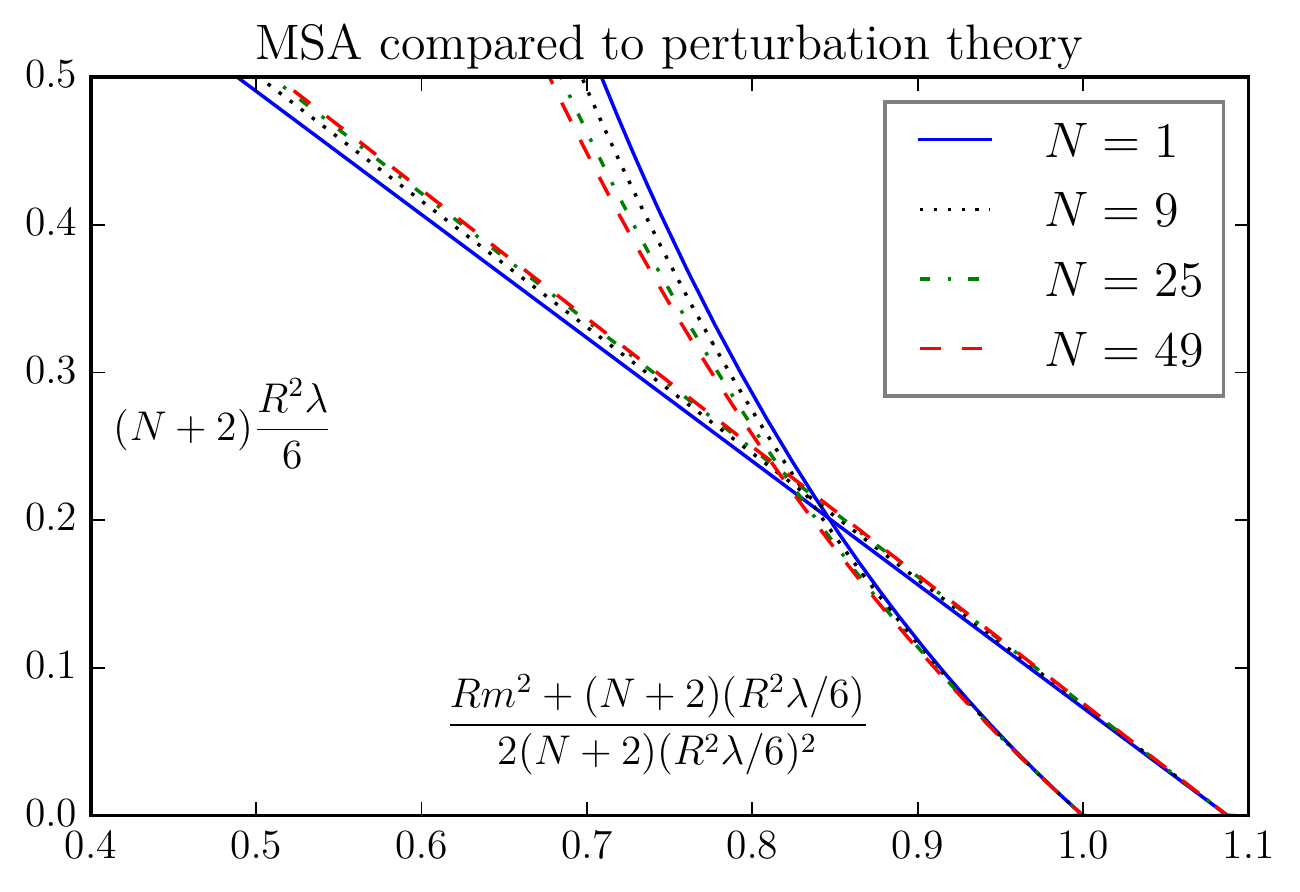}
\end{center}
\caption{\label{figure6}Detailed view of the small $\lambda$ behaviour of the critical line.
To better expose the differences we have subtracted the exact
first order term from $R m^2$, and further scaled the axes.  
Thus, the $\lambda^3$-term of the exact perturbation expansion
(\ref{PerturbationExpansionOfCriticalLine}) becomes a straight
line. It shows a weak dependence of the spin dimensionality $N$
beacuse the third order diagrams in figure~\ref{fig:SelfEnergyDiagrams}
are proportional to both $(N+2)^2$ and $(N+2)(N+8)$, cf.~\eqref{QFTWeakCouplingExpansion}.
The curved lines represent the MSA solution
starting at position $1.0$ for $\lambda=0$ (by construction).
The offset in starting point relative to the perturbative straight line is
due to an 8\,\% error in the MSA prediction of the $\lambda^2$-behaviour;
the differerence in the starting slope is due to a 20\,\% error in
the MSA prediction of the $\lambda^3$-behaviour, cf.~eq.~\eqref{SmallLambdaExpansion}.}
\end{figure}

\begin{figure}
\begin{center}
		\includegraphics[width=0.9 \textwidth]{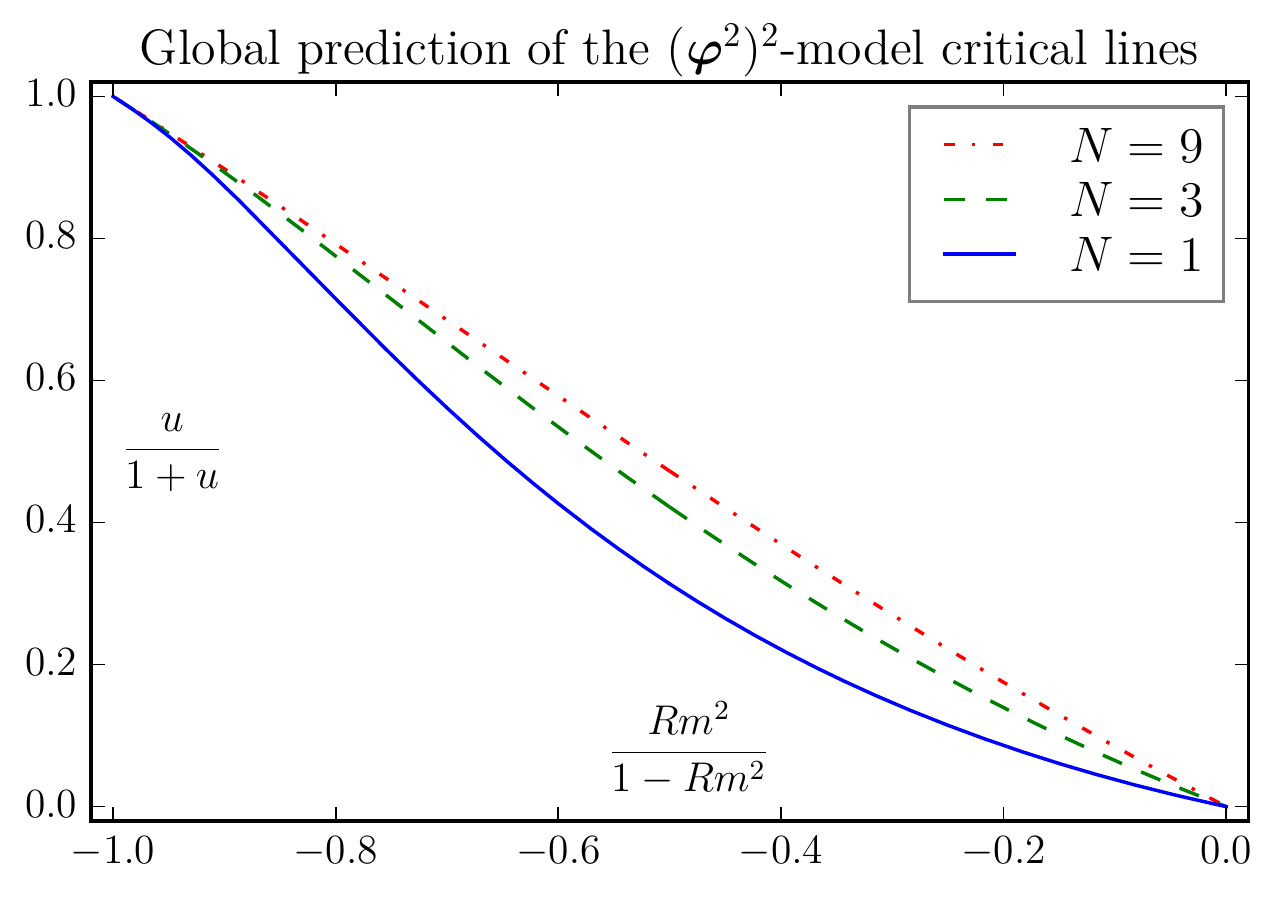}
\end{center}
\caption{\label{figure7}A global view of the the MSA prediction for the critical line.
To cover the full range of interaction strengths a non-linear parametrization of axes
is used, with $u=\frac{1}{6}N R^2 \lambda$. The critical line is linear in $\lambda$ for
both very small and very large $\lambda$, behaving respectively like
$Rm^2=-\frac{1}{6}(N+2) R^2\lambda + \mathcal{O}(\lambda^2)$ and
$R m^2 = -\frac{1}{6} N R^2\lambda -2/N + \mathcal{O}(\lambda^{-1})$,
cf.~eqs.~\eqref{SmallLambdaExpansion} and \eqref{LargeLambdaExpansion}.
The small-$\lambda$ behaviour is exact, the large-$\lambda$ behaviour
is about 3\,\% too high (for $\bm{\varphi}$-dimension $N=1$ and space-time
dimension $d=4$). This is expected to become more accurate as
$N$ increases.
}
\end{figure}

\section{Numerical computations}
\label{sec:MSAnumeric}

The integrals in eqs.~\eqref{MeanFieldCriticalLine} and \eqref{MSACriticalLine2} can be transformed
to standardized integrals, like $\mathcal{K}_{2n}(x)$ defined by \eqref{IntegralsKn}
when $m^2 + 2 d_{\text{eff}} > 0$,
and $\mathcal{I}_{2n}(x)$ defined by \eqref{IntegralsIn} when $m^2 + 2 d_{\text{eff}} < 0$.
For the mean field case,
eq.~\eqref{MeanFieldCriticalLine}, $d_{\text{eff}} \to d$. The $\mathcal{K}_{2n}(x)$ and $\mathcal{I}_{2n}(x)$ when $n$ is non-negative integer, can be expressed in terms of modified
Bessel functions of order $\frac{1}{4}$ and $\frac{3}{4}$,
and more elementary functions, cf.~appendix \ref{subsec:ExactExpressions}.

In terms of these functions the exact mean field and
MSA solutions for the critical line can be written
in parametric form
\begin{subequations}
\begin{equation}
\begin{split}
	\lambda &= 6 \left[\frac{2}{NR} \frac{\mathcal{K}_{2(n+1)}(x)}{\mathcal{K}_{2n}(x)}\right]^2,\\
	m^2 &= -1/(NR) + \sqrt{\lambda/6}\, x,
\end{split}
\end{equation}
with $0 < x < \infty$. The spin dimensionality is $N=2n+1$. This covers the region $m^2 + (1/R) > 0$, where $m^2 \to 0^-$ when $x\to\infty$.
The region $m^2 + (1/R) < 0$ is covered by the parametrization
\begin{equation}
\begin{split}
	\lambda &=  6 \left[ \frac{2}{NR}\frac{ \mathcal{I}_{2(n+1)}(y)}{\mathcal{I}_{2n}(y)}\right]^2, \\
	m^2 &= -1/(NR) - \sqrt{\lambda/6}\, y,
\end{split}
\end{equation}
with $0 < y < \infty$, where $m^2 \to -\infty$ when $y\to\infty$.
\end{subequations}
The $K_{2n}$ and $I_{2n}$  are integrals given by eqs.~\eqref{IntegralsKn}
and \eqref{IntegralsIn} in appendix \ref{subsec:ExactExpressions}.

A global view of the resulting critical lines is given in figure~\ref{figure7} where non-linear quantities in terms of $\lambda$ and $m^2$ are used. In figure~\ref{figure5} comparison with perturbation theory has been performed ($N=1$), and in figure~\ref{figure6} this has been performed in more detail.

With a linear scale in terms of $\lambda$ and $m^2$ the qualitative behavior of the global phase diagram follows in an obvious way from the term linear in each of eqs.~(\ref{SmallLambdaExpansion}) and (\ref{LargeLambdaExpansion}) and the constant term in the latter. The  slopes of the straight lines formed by these terms, will be the same in the limit $N\rightarrow\infty$ by which they join in just one straight line. If the $N\rightarrow\infty$ curve had been drawn on figure ~\ref{figure7}, it would also be a straight line there. For finite $N$ these two straight lines will intersect at $-m^2=(N+2)/(NR)$ (where  $R=a_1=0.154933\cdots$). Thus at this position the critical line will have an intermediate slope. So altogether the critical line starts with one slope at $\lambda=0$. Then the magnitude of this slope increases somewhat in a monotonic way towards its large $\lambda$ value (with $m^2$ on the horizontal axis). In view of the simplicity of this situation we have not drawn separate figures for these slightly bent lines.

%
%
%
%
%
%
%


\appendix

\section{Gaussian model with  finite temperature and chemical potential}
\label{GaussianModelFiniteTemperatureChemicalPotential}

Consider the gaussian path integral
\begin{equation}
\Xi = \int \text{e}^{-\int dt\,\left[(\partial_t -\bar{\mu}){\varphi}^* (\partial_t + \bar{\mu}){\varphi} 
	+\bar{\varepsilon}^2 \varphi^* \varphi\right]}\,\mathcal{D}\varphi^* \mathcal{D}\varphi,
\label{QFTPartitionFunction}
\end{equation}
where $\mu\equiv \hbar\bar{\mu}$ is a chemical potential, and $\varepsilon\equiv \hbar\bar{\varepsilon}$ a site energy.
The integration is over all complex continuous functions $\varphi(t)$ defined on a 
circle of circumference $\tau = \hbar\beta$. Expand $\varphi(t)$ in a Fourier series,
$\varphi(t) = \sum_n \varphi_n\,\text{e}^{-\text{i}\omega_n t}$, with Matsubara frequencies $\omega_n = 2\pi n/\tau$.
Define further the integration measure as $\mathcal{D}\varphi^* \mathcal{D}\varphi 
= \tau^{-2}d\varphi^*_0 d\varphi_0\prod_{n\ne0} \omega_n^2 d\varphi^*_n d\varphi_n$, where each complex
Fourier coefficient is integrated over the complex plane (normalized such that $\int \text{e}^{-z^* z}\,dz^* dz = 1$). This gives
\begin{align}
\Xi &= \int \text{e}^{-(\bar{\varepsilon}^2 - \bar{\mu}^2)\varphi_0^* \varphi_0}\,\tau^{-2} d\varphi_0^* d\varphi_0\prod_{n\ne0} \omega_n^2\;
\int\, \text{e}^{-\left[(\omega_n + \text{i}\bar\mu)^2 + {\bar\varepsilon}^2\right] \varphi^*_n \varphi_n}  d\varphi^*_n d\varphi_n\nonumber\\
&= \frac{(k_\text{B} T)^2}{{\varepsilon}^2-{\mu}^2} 
\prod_{n\ne0}{ \left(1 + \frac{2\text{i}\bar\mu}{\omega_n} + \frac{\bar{\varepsilon}^2 - \bar{\mu}^2}{\omega^2_n} \right)^{-1}}
= \frac{ \text{e}^{-\beta{\varepsilon}/2}}{1-\text{e}^{-\beta({\varepsilon}-{\mu})}}\,
\frac{ \text{e}^{-{\beta\varepsilon}/2}}{1-\text{e}^{-\beta({\varepsilon}+{\mu})}}.
\end{align}
The last equality follows (after some work) from the relation $\prod_{n=1}^\infty (1+ z^2/n^2) = \sinh \pi z/\pi z$.
This demonstrates that the path integral \eqref{QFTPartitionFunction} defines the grand canonical partition function for
non-interacting bosons (particles and antiparticles) occupying a quantum dot at temperature $T$. There is a zero-point energy $\frac{1}{2}\varepsilon$
per particle type, a site occupation energy $\varepsilon$, and a chemical potential $\mu$ for particles and $-\mu$ for antiparticles.
It should be clear that this connenection
can be extended to arbitrary many harmonic oscillators,
and --- by including factors like
$\text{exp}\left[\int dt\, \varphi^*(t)h(t)+ h^*(t)\varphi(t)\right]$ ---
to the corresponding generating functions.

\section{MSA solution of gaussian model}       
\label{secA}

The partition function of the gaussian model can also be found by using the lattice gas picture of a mixture of particles. Again we will use the MSA. With pressure $p$ we have with eq.~(\ref{1b}) 
\begin{equation}
\beta p=I+\sum_s\beta\mu_s\rho_s.
\label{71}
\end{equation}
With eqs.~(\ref{9b}) and (\ref{10a})
\begin{equation}
I=I_R+\frac{1}{2}{\m}^2 \tilde v(0)+I_1.
\label{72}
\end{equation}
Further from eq.~(\ref{17}) one has the effective chemical potentials for the reference system (the ones that give the resulting densities $\rho_s$)
\begin{equation}
\beta\mu_{es}=\ln z_{es}=\ln f_e(s)+\beta{\cal H}_e s.
\label{73}
\end{equation}
Now with eqs.~(\ref{6}), (\ref{18}), (\ref{3.5}), and (\ref{26a}) the reference system (free spins) pressure is
\begin{equation}
\beta p_R=I_R+\sum_s\beta\mu_{es}\rho_s=\ln\rho_0=\frac{1}{2}\ln R+\frac{1}{2}\beta {\cal H}_e {\m}
\label{74}
\end{equation}
in accordance with eq.~(\ref{5}). Finally with eqs.~(\ref{1}) and (\ref{73}) one has
\begin{equation}
\begin{split}
\Delta I&=\sum_s(\beta\mu_s\rho_s-\beta\mu_{es}\rho_s)\\
&=\sum_s \left(\frac{1}{2}c_0 s^2-\tilde v(\mathbf{0}) {\m}s\right)\rho_s
=\frac{1}{2}c_0(R+{\m}^2)-{\m}^2 \tilde v(\mathbf{0}),
\label{75}
\end{split}
\end{equation}
when eqs.~(\ref{MeanValue}) and (\ref{7c}) are used. Adding together we find
\begin{equation}
\begin{split}
\beta p&=\beta p_R+\Delta I+\frac{1}{2} {\m}^2
\tilde v(\mathbf{0})+I_1\\
&=\frac{1}{2}\ln R -
\frac{1}{2(2\pi)^d}\int_{-\pi}^\pi\cdots\int_{-\pi}^\pi\,
\ln[1-R\tilde v(\mathbf{0})]\, d\mathbf{k}+
\frac{1}{2}\beta\mathcal{H} {\m},
\label{76}
\end{split}
\end{equation}
when (\ref{18}) for $\beta{\cal H}_e$ is inserted. With
$R^{-1}=R_0^{-1}+c_0$, and 
\begin{equation*}
	\beta{\cal H}=\beta{\cal H}_e
	-\tilde v(\mathbf{0}) {\m}=[R^{-1}-\tilde v(0)]{\m},\quad
	\text{where }
	\tilde v(\mathbf{0})=c_0-\beta \tilde\psi(\mathbf{0}),
\end{equation*}
this gives precisely the gaussian model result (\ref{Pressure}).

\section{Integral representation of the lattice propagator}
\label{LatticePropagator}

We want to evaluate integrals of the form
\begin{equation}
G(\bm{x}; m^2, \nu) \equiv  \frac{1}{V_{\text{BZ}}} \int_{\text{BZ}} 
\frac{\text{e}^{\text{i}\bm{k}\cdot\bm{x}}}{\left(m^2 + 2d - \sum_{n=1}^d \cos k_n \right)^{\nu}}\, d\bm{k},
\label{c1}
\end{equation}
where the integral is over the Brillouin zone BZ of a $d$-dimensional hypercubic lattice,
with volume $V_{\text{BZ}}$. The lattice propagator is obtained for $\nu=1$. By use of the formula
$\int_0^\infty dt\, t^{\nu-1}\,\text{e}^{-a t} = \Gamma(\nu)\, a^{-\nu}$ this can be rewritten as
\begin{align}
G(\bm{x}; m^2, \nu) &= \frac{1}{\Gamma(\nu)} \int_0^\infty  \frac{dt}{t}\, t^\nu \,\text{e}^{-m^2 t } 
\prod_{n=1}^d e^{-2t}\,\int_{-\pi}^\pi \frac{dk_n}{2\pi}\, \text{e}^{2t\cos k_n}  \text{e}^{\text{i} k_n x_n}\nonumber\\
&=  \frac{1}{\Gamma(\nu)} \int_0^\infty  \frac{dt}{t}\, t^\nu \,\text{e}^{-m^2 t } 
\,\prod_{n=1}^d  \text{e}^{-2t} I_{x_n}(2t),
\label{c2}
\end{align}
where $I_n(t)$ is a modified Bessel function.

\section{Evaluation of integrals and solution of the closure relation}
\label{sec:EvaluatingR}

The results presented in sections \ref{sec:MSAanalytic} and  \ref{sec:MSAnumeric}
requires evaluation of 
$\langle \bm{\varphi}^2 \rangle$. In this appendix we give some details of how it
can be evaluated in various parameter ranges, and how the closure relation
\begin{equation}
	N R = \langle \bm{\varphi}^2 \rangle
	\label{MSACoreCondition}
\end{equation}
can be solved.
A useful starting point is the formula 
\begin{equation}
	\langle \bm{\varphi}^2 \rangle = 
	\frac{d}{da} \ln \int \bm{e}^{a \bm{\varphi}^2 -\frac{1}{4!}\lambda (\bm{\varphi}^2)^2}  d\bm{\varphi} \equiv \frac{d}{da} \ln Z,
	\label{OriginalFormula}
\end{equation}
where $a = -\frac{1}{2}(m^2 + R^{-1})$. The algebraic manipulations can
be performed by computer algebra, and verified against low order manual
calculations.

\subsection{Perturbation expansion for small $\lambda$}
\label{subsec:SmallLambda}

We use the fact that $\frac{d}{da} \ln Z = 2 R^2 \frac{d}{dR} \ln Z$ when $a  = -\frac{1}{2}(m^2 + R^{-1})$, and
transform eq.~\eqref{OriginalFormula} to the relation
\begin{equation*}
\begin{split}
   \langle \bm{\varphi}^2 \rangle &=  N R \\
   &+ 2 R^2 \frac{d}{dR} \ln \int_0^\infty  
   t^{(N-2)/2}\, \text{e}^{-t}\,\text{e}^{-m^2 R t - \frac{1}{6}\lambda R^2 t^2} dt,
\end{split}
\end{equation*}
which is satisfied for $\lambda=0$ if we assume $m^2=0$ to $0^\text{th}$ order in $\lambda$.
We thus make the ansatz that $m^2 = \sum_{n=1}^\infty m^2_n \lambda^n$, expand the integral
and its logarithm in powers of $\lambda$, and insert this expansion into eq.~\eqref{MSACoreCondition}.
The resulting equation can now be solved recursively for the coefficients $m^2_n,\;n=1, 2, \ldots$. The first
terms of this expansion are given in eq.~\eqref{SmallLambdaExpansion}.

\subsection{Asymptotic expansion for large $\lambda$}
\label{subsec:LargeLambda}
A first estimate of the integral in eq.~\eqref{OriginalFormula} by
the Laplace method gives $\ln Z \approx 6a^2/\lambda$,
implying that $a \approx \frac{1}{12} NR \lambda$ for large $\lambda$.
Hence the parameter $v\equiv N R\lambda/(12a)$ is close to 1 in this
limit, with a deviation which can be expanded in the small quantity
$\epsilon \equiv 1/(NRa)$. With
\begin{equation*} 
	\varepsilon \equiv  \sqrt{{\lambda}/{12}}\, a^{-1} 
	= \sqrt{v\, \epsilon},
\end{equation*}
eq.~\eqref{OriginalFormula} can be transformed to the form
\begin{equation}
\begin{split}
	v &= 1 + \text{$\frac{1}{2}$}(N-2) v\epsilon\\
	&- v\epsilon \Big[ \varepsilon \frac{d}{d\varepsilon}
	\int^\infty_{-1/\varepsilon} (1+\varepsilon\, t)^{(N-2)/2}\,\text{e}^{-\frac{1}{2} t^2} dt
	\Big]_{\varepsilon^2 = v \epsilon}.
	\label{Large_Lambda_Expansion}
\end{split}
\end{equation}
An asymptotic expansion of the integral as $\varepsilon\to0^+$ is straightforward to generate by binomial expansion and term-by-term integration. eq.~\eqref{Large_Lambda_Expansion} may
next be solved as an expansion,
$v = 1 + \sum_{n\ge1} v_n\,\epsilon^n$,
such that one finds (from the definition of $v\equiv\epsilon/\delta$),
\begin{equation}
	{12}/(N^2 R^2 \lambda) \equiv \delta 
	= \epsilon\,
	\Big(1 + \sum_{n\ge1} v_n \epsilon^n \Big)^{-1}.
	\label{delta_expansion}
\end{equation}
This relation can be inverted to an expansion
$\epsilon = \delta\,(1 + \sum_{n\ge1} \epsilon_n \delta^n)$.
It finally follows that
\begin{equation}
    N R m^2 = -N - (2/\delta)
    \Big(1 + \sum_{n\ge1} \epsilon_n \delta^n \Big)^{-1}.
\end{equation}
The first few terms of this expansion
are given in eq.~\eqref{LargeLambdaExpansion}.

\subsection{Asymptotic expansion for large $N$}
\label{subsec:LargeN}

Eq.~\eqref{MSACriticalLine2} can be written as
\begin{equation}
\tilde{R} = \frac{\int_0^\infty \rho\,
	\text{e}^{\frac{1}{2}(N-2)\left[\ln \rho + 2 a \rho 
		- \frac{1}{2}\tilde{u} \rho^2 \right]} d\rho}{
	\int_0^\infty \text{e}^{\frac{1}{2}(N-2)\left[\ln \rho + 2 a \rho 
		- \frac{1}{2}\tilde{u} \rho^2\right]} d\rho} \equiv \langle \rho \rangle,
\label{largeN_MSA}
\end{equation}
where we have introduced $\rho = \bm{\varphi}^2/(N-2)$,
$\tilde{R} = NR/(N-2)$, and $\tilde{u}=\frac{1}{6}(N-2)\lambda$.
When $(N-2)$ is large the main contribution to the
integrals comes from the vicinity of the maximum point $\rho_0$, satisfying
\begin{equation}
	a = -\frac{1}{2}\left(\rho_0^{-1} - \tilde{u} \rho_0\right).
	\label{MassRelation}
\end{equation}
Hence, to leading order
in $N$ eq.~\eqref{largeN_MSA} becomes $\tilde{R} = \langle \rho \rangle = \rho_0$.
I.e.,
\begin{equation*}
\begin{split}
   a & \equiv -\frac{1}{2} \left(m^2 + R^{-1}\right) = 
   -\frac{1}{2}\left(\tilde{R}^{-1} -\tilde{u}\,\tilde{R}\right),\\
 & \text{or}\quad m^2 = - \tilde{u}\,\tilde{R} = -\frac{1}{6} NR\lambda,
\end{split}
\end{equation*}
since $\tilde{R} = R$ to leading order.
To proceed we expand the
integrals in \eqref{largeN_MSA} in powers of
$\bar{\nu} \equiv 1/(N-2)$. By introducing $\bar{v} \equiv \rho_0^2 \, \tilde{u} = 
\frac{1}{6}(N-2)\rho_0^2 \, \lambda$, eq.~\eqref{largeN_MSA} acquires the form
\begin{equation}
	\tilde{R} = 
	\rho_0\,\Big[1 + \sum_{k\ge 1} r_k(\bar{v})\; \bar{\nu}^{k}\Big],
	\label{Nexpansion1}
\end{equation}
where the coefficients $r_k$ are
rational functions in $\bar{v}$. Eq.~\eqref{Nexpansion1} can be viewed as
an equation for $\rho_0$,
or equivalently for $\bar{v}$. We solve it by introducing $\bar{u} \equiv \tilde{R}^2\,\tilde{u}$,
such that $\tilde{R}/\rho_0 = \sqrt{\bar{u}/\bar{v}}$. Then eq.~\eqref{Nexpansion1}
may be rewritten as
\begin{equation}
	\bar{u} = \bar{v}\,\Big[1 + \sum_{k\ge1} r_k(\bar{v})\;\bar{\nu}^{k} \Big]^{2},
\end{equation}
and solved iteratively for $\bar{v}$ order by order in $\bar{\nu}$, leading to
$\bar{v}=\bar{u}(1 + \sum_{k\ge1} s_k(\bar{u})\,\bar{\nu}^k)$.
Finally, using \eqref{MassRelation}, 
the expression for the critical line can be expressed as
\begin{equation}
	\tilde{R}\,m^2 = -\tilde{R} R^{-1} + \sqrt{\bar{u}/\bar{v}}
	-\sqrt{\bar{u}\bar{v}}.
\end{equation}
With the expansion for $\bar{v}$ known, this is straightforward
to expand once more. Finally, for simpler comparison with the expressions for small
and large $\lambda$, we rewrite the expression as an expansion in 
$\nu \equiv 1/N = \bar{\nu}/(1+2\bar{\nu})$ and 
$u \equiv \frac{1}{6} N R^2 \lambda = \bar{u}/(1+2\bar{\nu})$.  
The first terms of this expansion are given
in eq.~\eqref{OneOverNexpansion}.

\subsection{Exact expressions}
\label{subsec:ExactExpressions}

The integrals in eq.~\eqref{MeanFieldCriticalLine} can be
evaluated exactly. The results involve different functions
depending on whether $N$ is even or odd. We will here only
discuss the more complicated cases when $N=2n+1$ is odd.
By introducing a new variable of integration,
$t^4=\lambda(\bm{\varphi}^2)^2/4!$, the relevant integrals
can be transformed the form
\begin{equation}
	\mathcal{K}_{2n}(u) \equiv 2 \int_{0}^{\infty}  t^{2n}\, \text{e}^{-t^4-u\,t^2} dt,
	\label{IntegralsKn}
\end{equation}
and the corresponding expression with $u$ replaced by $-v$,
\begin{equation}
	\mathcal{I}_{2n}(v) \equiv 2 \int_0^{\infty}  t^{2n}\, \text{e}^{-t^4+v\,t^2} dt.
	\label{IntegralsIn}
\end{equation}
The integral~\eqref{IntegralsKn} can be found in
the integral table  \cite{GradshteynRyzhik81} for $n=0$.
Then higher values of $n$ by can be derived by
repeated differentation with respect to $u$.
The corresponding integrals~\eqref{IntegralsIn} can next be found by analytic continuation, 
$u \to u\, \text{e}^{\text{i}\phi}$ with $0 \le \phi \le \pi$. Introduce
\begin{equation}
\begin{split}
     f(u) & \equiv  u^{1/2}\, \text{e}^{\frac{1}{8} u^2}\, K_{1/4}({\textstyle \frac{1}{8}} u^2),\\
     g(u) &\equiv  u^{3/2}\, \text{e}^{\frac{1}{8} u^2}\, K_{3/4}({\textstyle \frac{1}{8}} u^2).
\end{split}
\end{equation}
We find that $\mathcal{K}_0(u) = \frac{1}{2} f(u)$. From the recursion relations for Bessel functions \cite{AbramowitzStegun}, specifically $K'_{1/4}(z) = -\frac{1}{4z}K_{1/4}(z) - K_{3/4}(z)$ and $K'_{3/4}(z)= -K_{1/4}(z) - \frac{3}{4z}K_{3/4}(z)$, it next follows that
\begin{equation}
    \frac{d}{du} \begin{pmatrix} f(u)\\ g(u) \end{pmatrix} = 
    \frac{1}{4}\begin{pmatrix} u & -1\\ -u^2 & u\end{pmatrix} 
    \begin{pmatrix} f(u)\\ g(u) \end{pmatrix}.
\end{equation}
Hence we may write
\begin{equation}
      \mathcal{K}_{2n}(u) = P_n(u) f(u) + Q_n(u) g(u),
      \label{IntegralsKnForm}
\end{equation}
where $P_n$ and $Q_n$ are polynomials in $u$, satisfying the recursion relation
\begin{equation}
    \begin{pmatrix} P_{n+1}(u) \\ Q_{n+1}(u) \end{pmatrix} 
    = -\frac{d}{du}  \begin{pmatrix} P_{n}(u) \\ Q_{n}(u) \end{pmatrix} 
    	- \frac{1}{4}\begin{pmatrix} u & -u^2\\-1 & u\end{pmatrix}
    	\begin{pmatrix} P_{n}(u) \\ Q_{n}(u) \end{pmatrix},
\end{equation}
starting with  $P_0 = \frac{1}{2}$, $Q_0 = 0$. The next few polynomials are listed in table \ref{tab:PolynomialsPQ}.
Analytic continuation leads to the representation

\begin{table}[h]
\begin{center}
\begin{tabular}{|l|c|c|}
\hline
&&\\[-2.4ex]
$n$& $(-)^n\, 2^{n+2}\, P_n$ & $(-)^{n+1}\, 2^{n+2}\, Q_n$\\[0.1ex] 
\hline
&&\\[-2.3ex]
$0$&$2$ & $0$\\
$1$&$u$ & $1$\\
$2$&$u^2 + 2$ & $u$\\
$3$&$u^3 + 5 u$ & $u^2 + 3$\\
$4$&$u^4 + 10 u^2 + 10$ & $u^3 + 8 u$\\
$5$ & $u^5 + 17 u^3 + 45 u$ & $u^4 + 15 u^2 + 21$ \\[0.3ex]
\hline
\end{tabular}
\end{center}
\caption{\label{tab:PolynomialsPQ} The first few polynomials $P_n$ and $Q_n$ in
\eqref{IntegralsKnForm} and \eqref{IntegralsInForm}.}
\end{table}

\begin{equation}
	\mathcal{I}_{2n}(v) = {P}_n(-v) \bar{f}(v) + {Q}_n(-v) \bar{g}(v),
	\label{IntegralsInForm}
\end{equation}
where
\begin{subequations}
\begin{align}
	\bar{f}(v) &= v^{1/2} \text{e}^{\frac{1}{8} v^2} \left[ K_{1/4}({\textstyle\frac{1}{8}} v^2) 
    	+ \sqrt{2}\,\pi I_{1/4}({\textstyle \frac{1}{8}} v^2)\right],\\
	\bar{g}(v) &= v^{3/2} \text{e}^{\frac{1}{8} v^2} \left[ K_{3/4}({\textstyle\frac{1}{8}} v^2) 
    	+ \sqrt{2}\,\pi I_{3/4}({\textstyle \frac{1}{8}} v^2)\right].    	
\end{align}
\end{subequations}

The final analytic expressions have been checked (successfully) against direct numerical
evaluation for a number of cases.

\bibliographystyle{JHEP}
\bibliography{BibPaper0}

\end{document}